\documentclass[11pt,a4paper]{article}

\usepackage[utf8]{inputenc}
\usepackage[T1]{fontenc}
\usepackage{mathptmx}
\usepackage[scaled=0.92]{helvet}
\usepackage{amsmath,amssymb}
\usepackage{bm}
\usepackage{graphicx}
\usepackage{booktabs}
\usepackage{array}
\usepackage[margin=2.5cm]{geometry}
\usepackage[font=small,labelfont=bf,labelsep=period]{caption}
\usepackage{xcolor}
\usepackage[authoryear,round,sort]{natbib}
\usepackage{xurl}
\usepackage[colorlinks=true,linkcolor=blue!50!black,citecolor=blue!50!black,urlcolor=blue!50!black]{hyperref}

\setcitestyle{aysep={,},yysep={,},notesep={, }}
\newcommand{\um}{\,\mu\mathrm{m}}
\newcommand{\Cint}{C_{A,\cap}}
\newcommand{\phiacc}{\phi_{\mathrm{acc}}}
\newcommand{\RMST}{\mathrm{RMST}}

\title{Platelet-driven contraction enlarges accessible space in layered fibrin networks without uniformly accelerating particle escape}

\author{Juan P. Staforelli-Vivanco \and Paolo Teppa Gutierrez\thanks{Corresponding author: \href{mailto:pteppa2018@udec.cl}{pteppa2018@udec.cl}}\\[6pt]
\normalsize\normalfont Departamento de F\'isica, Facultad de Ciencias F\'isicas y Matem\'aticas,\\
\normalsize\normalfont Universidad de Concepci\'on, Concepci\'on, Chile}

\date{}

\begin{document}
\maketitle

\begin{abstract}
Platelet traction contracts fibrin networks, yet whether contraction promotes or hinders the escape of suspended particles is unclear. We coupled an overdamped Langevin model of platelet--fibrin mechanics to Brownian transport of finite-size tracers through frozen network states. Nine mechanical replicas were evolved at four filopodial traction baselines (0--8.8\,nN), stacked into 5-$\mu$m-thick domains of 50 layers and probed at 0--15\,s with 0.5-$\mu$m tracers (29,952 trajectories, 100\,s observation). Benchmarked against experimental clot-contraction kinetics, the simulated contraction is post-lag, within one order of magnitude of experimental rates and corresponds to the onset of the contraction phase. Relative to the matched zero-baseline control, every active-baseline state showed greater projected contraction and a larger tracer-accessible volume ($+9.3\%$ to $+27.6\%$; exact sign test $p=0.0039$ per condition). At 15\,s, mean survival at 100\,s fell from 27.3\% to 14.3\% and the restricted mean exit time from 56.5 to 44.0\,s between the lowest and highest baseline, and all nine replicas shortened residence at 8.8\,nN. At lower baselines, responses differed among replicas, and nearly all retained tracers with a classified start were connected to an exit. Contraction therefore enlarges accessible space without uniformly accelerating escape, separating geometric accommodation from transport kinetics.
\end{abstract}

\noindent\textbf{Keywords:} blood clot contraction; fibrin network; platelets; first-passage time; Brownian dynamics; porous media; optical speckle; neonatal haemostasis

\bigskip

\section{Introduction}

Blood clots are not static plugs. Within minutes of formation, activated platelets pull on the fibrin network through filopodia, compacting the clot and expelling serum in a process known as clot contraction or retraction \citep{litvinov2023,cines2014,tutwiler2016}. Contraction changes clot size, stiffness and permeability, and its impairment has been associated with thrombotic disease \citep{litvinov2023}. At the scale of individual platelets, traction forces in the nanonewton range bend and bundle fibres and reorganize the local architecture \citep{lam2011,kim2017,michael2023}.

How this reorganization affects transport inside the clot is less clear. Clots retain and release cell-derived microparticles, extracellular vesicles, bacteria and therapeutic nanocarriers, whose sizes are comparable to the pores of the fibrin network \citep{zubairova2015,weisel2017}. For such particles, the relevant question is not only how much void space remains but whether that space is accessible to a particle of finite size, whether it is connected to an exit, and how long a particle needs to find that exit. Projected-area contraction, which is the standard experimental readout \citep{tutwiler2016,prakhya2022}, does not distinguish between these possibilities. Permeability and hydraulic tortuosity describe fluid flow \citep{wufsus2013,duda2011}, whereas the residence of a diffusing particle is a first-passage problem governed by geometry, connectivity and boundary conditions \citep{redner2001,benichou2014,holcman2014}. In crowded and porous media, these factors can produce slow, strongly heterogeneous escape even when the accessible volume fraction is substantial \citep{hofling2013,torquato2002}.

Mechanistic clot models have so far focused either on network mechanics and contraction \citep{kim2017,michael2023} or on enzyme transport during fibrinolysis \citep{bannish2014}, but the link between platelet-driven reorganization and the retention of finite-size particles has not been quantified. Here we couple a stochastic platelet--fibrin mechanics model, adapted from \citet{michael2023}, to reflecting Brownian transport of finite-size tracers. We ask three questions. (i) Does increasing platelet traction increase or decrease the volume accessible to a 0.5-$\mu$m particle? (ii) Does it change the survival and restricted mean exit time of centrally released particles? (iii) Can changes in retention be attributed to loss of connectivity to an exit or to slower motion within connected space?

We address these questions with nine independent mechanical replicas at four traction baselines, analysed at matched mechanical times and compared within replicas. To place the analysed states on an experimental time course, we benchmark the simulated contraction against published clot-contraction kinetics \citep{prakhya2022}. We find that traction consistently enlarges the accessible space but that escape kinetics respond heterogeneously across realizations, with a robust reduction in residence only at the highest baseline. The results show that accessible volume, availability of an exit route and the time required to follow it are distinct descriptors of clot transport.

\section{Results}

\subsection{A layered platelet--fibrin model with matched frozen states}

\begin{figure}[t]
  \centering
  \includegraphics[width=0.58\textwidth]{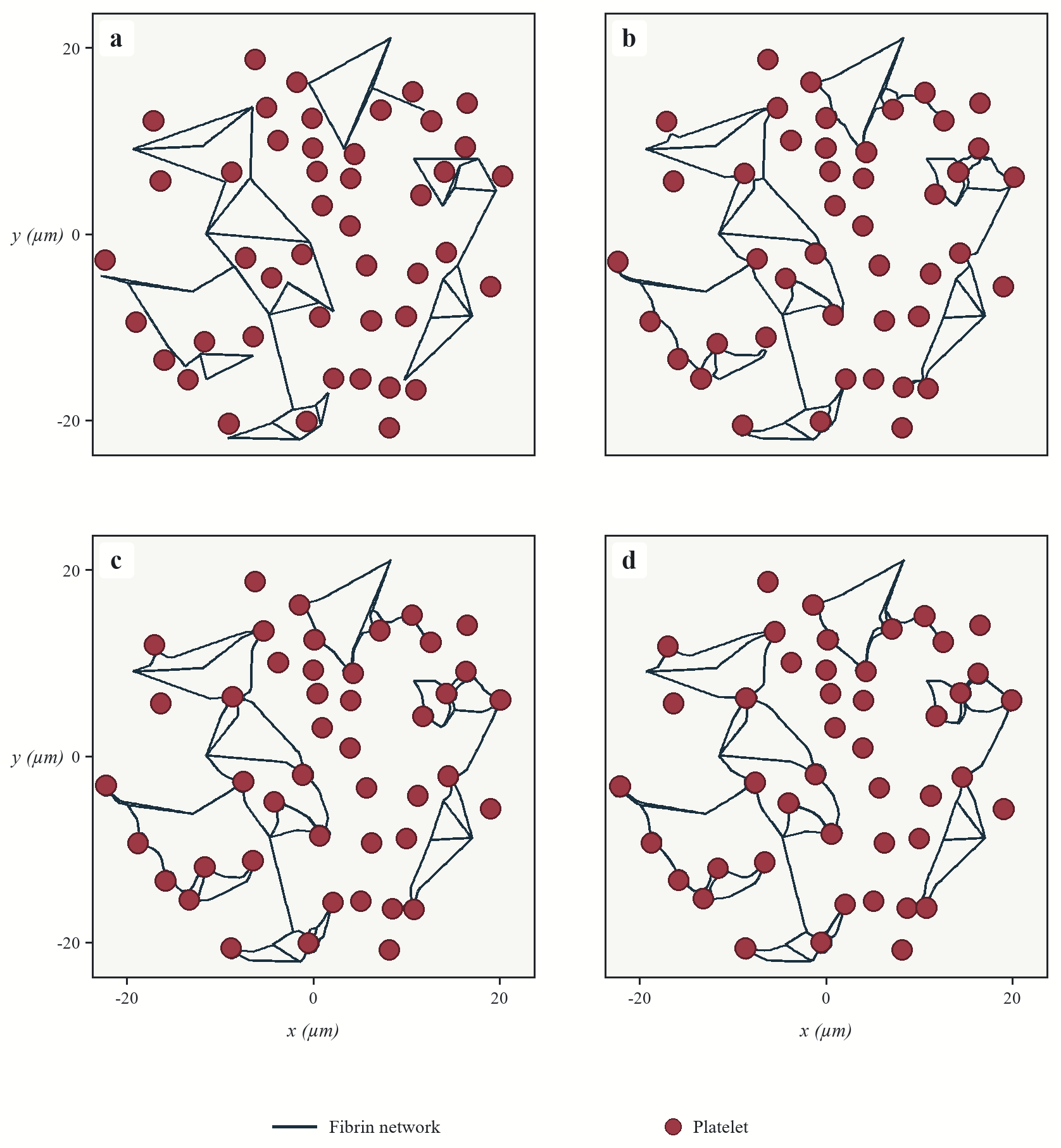}
  \caption{\textbf{Platelet traction reorganizes a single fibrin layer.} Representative layer shown at (a) 0\,s, (b) 2\,s, (c) 5\,s and (d) 10\,s of mechanical evolution ($f_0=2.233$\,nN). Disks, platelets ($r_p=1.13\um$); lines, fibrin backbone and cohesive links. All panels share spatial limits and scale. This layer illustrates the mechanics and is not part of the analysed cohort.}
  \label{fig:layer}
\end{figure}

\begin{figure}[t]
  \centering
  \includegraphics[width=0.5\textwidth]{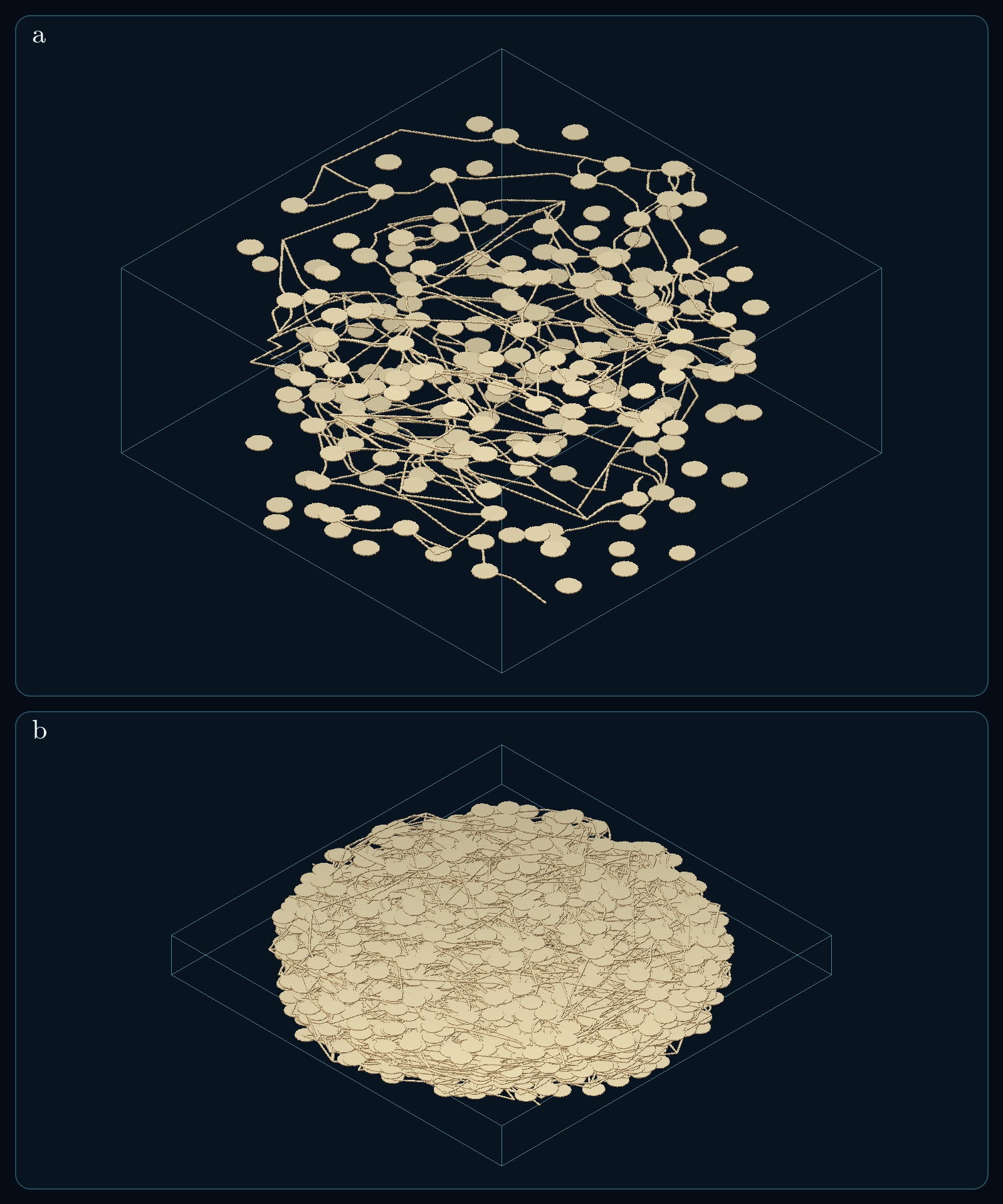}
  \caption{\textbf{Assembly of independent layers into the transport domain.} Replica 1 at $f_0=2.2$\,nN and $t_m=10$\,s. (a) Layers 1, 13, 25, 37 and 50 with separation expanded for display. (b) All 50 layers at $H=100$\,nm ($L_z=5\um$) without axial exaggeration. Fibrin, cohesive links and platelets share one colour. Voxels approximate the continuous solids for display only; transport uses the continuous primitives. The full lateral field is shown without clipping to the common analysed region.}
  \label{fig:stack}
\end{figure}

Each layer is a two-dimensional network of fibrin fibres and 45 platelets in a $47.6\times47.6\um^2$ domain, evolved with overdamped Langevin dynamics that include worm-like-chain fibre stretching, bending, fibre--fibre cohesion, platelet--fibrin contact and stiffness-dependent filopodial traction (Section~\ref{sec:methods}). Platelets bend and gather nearby fibres during the first seconds of evolution (Fig.~\ref{fig:layer}). Fifty independently generated layers of thickness $H=100$\,nm were stacked into a domain of thickness $L_z=5\um$ (Fig.~\ref{fig:stack}); layers are not mechanically coupled, so the stack approximates a multilayer material rather than a reconstructed three-dimensional clot.

Nine mechanical replicas were generated, each evolved at four filopodial baseline forces, $f_0\in\{0,2.2,4.4,8.8\}$\,nN, from a shared initial state. The zero baseline retains the stiffness-dependent part of the traction law and is therefore a low-traction control, not an inactive-platelet condition. Network states frozen at mechanical times $t_m=0$, 5, 10 and 15\,s provide 117 unique geometries. In each, 256 noninteracting tracers of radius $a_{\mathrm{tr}}=0.25\um$ were released in a central cylinder (Fig.~\ref{fig:S1}), reflected at solid surfaces and at the lateral boundary, and absorbed at either axial face. Trajectories were followed for $T_c=100$\,s, giving 29,952 histories. Survival $S(t)$ and the restricted mean survival time $\RMST(T_c)$, the mean of $\min(\tau,T_c)$ over all released tracers, were the primary transport readouts \citep{redner2001,royston2013}. Every active-baseline state was compared with its own zero-baseline control at the same mechanical time.

\subsection{Traction consistently increases contraction and accessible volume}

\begin{figure}[t]
  \centering
  \includegraphics[width=0.48\textwidth]{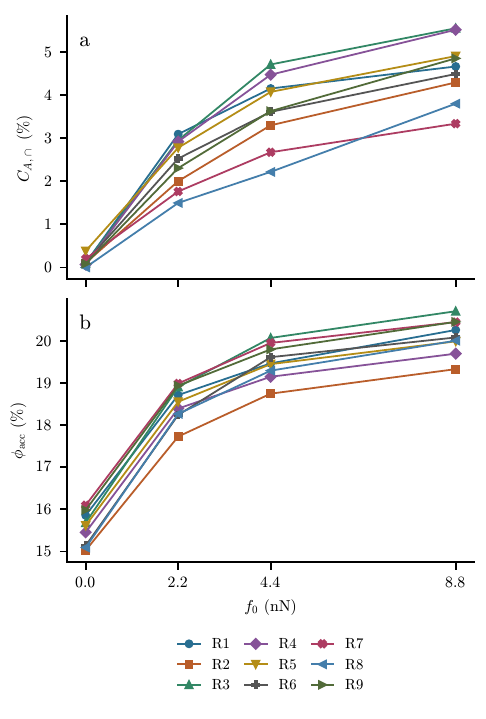}
  \caption{\textbf{Traction increases contraction and tracer accessibility in every replica.} Frozen geometry at $t_m=15$\,s. (a) Contraction of the common analysed area, $\Cint$. (b) Fraction of the domain accessible to tracer centres, $\phiacc$, for a tracer radius of $0.25\um$. Symbols identify the nine paired mechanical replicas (R1--R9). Lines guide the eye; no dose--response law is fitted.}
  \label{fig:access}
\end{figure}

In all 81 active-baseline versus control contrasts (nine replicas $\times$ three baselines $\times$ three mechanical times), both the contraction of the common analysed area, $\Cint$, and the fraction of the domain accessible to tracer centres, $\phiacc$, increased (Table~\ref{tab:transport}, Fig.~\ref{fig:access}). Treating replicas as the independent unit, the exact two-sided sign test gives $p=0.0039$ in each of the nine baseline--time cells ($p=0.035$ after Holm correction over the nine cells). Absolute accessible volume increased by 9.29--27.58\% relative to the matched control, so the rise in $\phiacc$ is not merely a consequence of a shrinking reference area. At 15\,s, mean $\Cint$ rose from 0.128\% at $f_0=0$ to 4.602\% at 8.8\,nN, and mean $\phiacc$ from 15.55\% to 20.12\%. Thus, in this model, contraction concentrates solid material and opens space accessible to particles of 0.5\,$\mu$m diameter.

\begin{table}[t]
\centering
\caption{\textbf{Contraction, accessibility and escape for frozen mechanical states.} Values are means over nine paired mechanical replicas; bracketed transport values are minimum--maximum across replicas, not confidence intervals. Each geometry has 256 tracer histories (diameter 0.50\,$\mu$m, $D=0.4\um^2\,\mathrm{s}^{-1}$, $T_c=100$\,s). The initial state is shared by all baselines. Contraction refers to the common analysed area.}
\label{tab:transport}
\small
\begin{tabular}{cccccc}
\toprule
$t_m$ (s) & $f_0$ (nN) & $\Cint$ (\%) & $\phiacc$ (\%) & $S(100)$ (\%) & $\RMST(100)$ (s)\\
\midrule
0  & all & 0.000 & 15.322 & 24.22 [12.11, 53.12] & 55.97 [43.53, 76.40]\\
\midrule
5  & 0   & 0.034 & 15.375 & 25.48 [10.55, 55.86] & 56.23 [42.51, 76.58]\\
5  & 2.2 & 1.112 & 17.180 & 21.35 [10.94, 41.02] & 51.63 [41.55, 69.39]\\
5  & 4.4 & 1.725 & 17.929 & 17.53 [4.69, 29.30]  & 50.04 [40.04, 61.34]\\
5  & 8.8 & 2.466 & 18.610 & 18.01 [9.77, 27.73]  & 48.74 [36.76, 62.34]\\
\midrule
10 & 0   & 0.064 & 15.436 & 27.08 [14.06, 60.55] & 56.91 [44.65, 78.36]\\
10 & 2.2 & 1.845 & 17.979 & 19.92 [12.11, 29.69] & 50.56 [41.76, 61.81]\\
10 & 4.4 & 2.727 & 18.806 & 18.45 [8.20, 34.77]  & 48.82 [36.58, 66.03]\\
10 & 8.8 & 3.586 & 19.517 & 15.54 [6.25, 29.69]  & 45.57 [33.20, 57.33]\\
\midrule
15 & 0   & 0.128 & 15.548 & 27.34 [17.19, 55.86] & 56.47 [45.46, 76.39]\\
15 & 2.2 & 2.426 & 18.531 & 18.53 [9.38, 32.81]  & 49.69 [38.32, 66.49]\\
15 & 4.4 & 3.649 & 19.512 & 16.97 [7.81, 39.06]  & 47.46 [35.54, 67.27]\\
15 & 8.8 & 4.602 & 20.116 & 14.32 [6.25, 29.69]  & 43.97 [32.27, 59.03]\\
\bottomrule
\end{tabular}
\end{table}

\subsection{Simulated contraction is consistent with the onset of experimental clot contraction}

\begin{figure}[t]
  \centering
  \includegraphics[width=\textwidth]{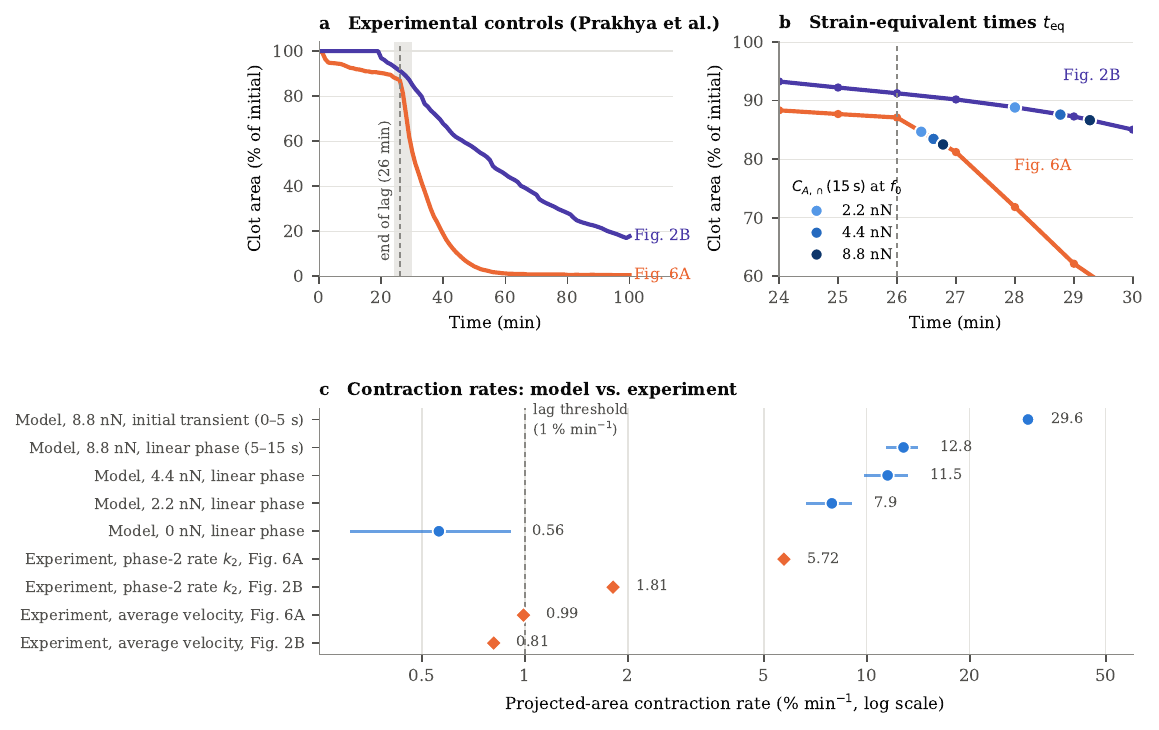}
  \caption{\textbf{Simulated contraction benchmarked against experimental clot-contraction kinetics.} (a) Control curves of \citet{prakhya2022} under standard conditions, digitized from their Figs.~2B and 6A; the dashed line marks the end of the lag (26\,min) and the shaded band the region enlarged in b. (b) Strain-equivalent times: points mark where each experimental curve has contracted by the replica-mean $\Cint(15\,\mathrm{s})$ beyond its value at the end of the lag, for the three active baselines. (c) Projected-area contraction rates. Circles, model (bars, 95\% bootstrap intervals over replicas); diamonds, experiment. Dashed line, lag threshold of \citet{prakhya2022}.}
  \label{fig:benchmark}
\end{figure}

To test whether the simulated mechanics produce contraction on an experimentally meaningful scale, we benchmarked them against the projected-area kinetics of washed-platelet clots reported by \citet{prakhya2022}, whose observable has the same normalization as $\Cint$. Their two control curves obtained under standard conditions were digitized and analysed with their own lag and phase definitions (Section~\ref{sec:benchmark}); they share a lag of 26\,min but differ about threefold in phase-2 rate (Table~\ref{tab:benchmark}, Fig.~\ref{fig:benchmark}a). In the overdamped model, multiplying all drag coefficients by a common factor reproduces the same sequence of network configurations on a proportionally stretched time axis (Section~\ref{sec:timescale}). Model time is therefore not calibrated physical time, so the comparison was made in strain and rate, and the factor relating simulated to experimental time is an output of the benchmark rather than an assumption.

Three observations are consistent with experiment. First, by the experimental lag criterion (contraction faster than 1\%\,min$^{-1}$), all 27 active-baseline replica states are post-lag from $t_m=0$, whereas eight of the nine zero-baseline replicas (mean 0.56\%\,min$^{-1}$) remain lag-like, as expected from the $\sim$130-s filopodial capture time at $f_0=0$ (Section~\ref{sec:timescale}). Second, without fitting any parameter to these data, the simulated linear-phase rates (7.9--12.8\%\,min$^{-1}$) are within one order of magnitude of the experimental phase-2 rates ($k_2=1.8$ and 5.7\%\,min$^{-1}$), exceeding them by $\lambda_r=1.4$--7.1 (Fig.~\ref{fig:benchmark}c); equivalently, multiplying all drag coefficients by $\lambda_r$ would bring the simulated rates onto the experimental phase-2 rates without changing any network configuration. Third, lowering $f_0$ from 8.8 to 2.2\,nN reduced the simulated rate to 0.61 of its maximum and removing the baseline reduced it to 0.05, bracketing the approximately threefold reduction of $k_2$ produced by the myosin inhibitor blebbistatin \citep{prakhya2022}.

In strain, the 15-s states cover a fraction $p=0.024$--0.056 of the experimental contraction extent, equivalent to 0.4--3.3\,min after the lag (Table~\ref{tab:benchmark}, Fig.~\ref{fig:benchmark}b). The transport analysis therefore concerns the onset of the contraction phase. Strain matching gives $\lambda_t=1.7$--13.1, larger than $\lambda_r$ for every baseline and both reference curves: simulated contraction starts at its maximal rate and decelerates, whereas the experimental curves accelerate after the lag.

\begin{table}[t]
\centering
\caption{\textbf{Benchmark of simulated contraction against experimental clot-contraction kinetics.} Top: control curves of \citet{prakhya2022} under standard conditions (washed mouse platelets, $5\times10^{7}$\,mL$^{-1}$; fibrinogen 250\,$\mu$g\,mL$^{-1}$; thrombin 20\,mU\,mL$^{-1}$), digitized from their Figs.~2B and 6A and analysed with their protocol. Bottom: simulation, means over nine replicas with 95\% bootstrap intervals over replicas for $r_{\mathrm{sim}}$; paired entries give values against the Fig.~6A\,/\,Fig.~2B reference. Post-lag, replicas with contraction faster than 1\%\,min$^{-1}$; $p$, fraction of the experimental extent reached at $t_m=15$\,s; $t_{\mathrm{eq}}$, post-lag time at which the experiment reaches the same contraction; $\lambda_t=t_{\mathrm{eq}}/15$\,s; $\lambda_r=r_{\mathrm{sim}}/k_2$; $r_{\mathrm{rel}}=r_{\mathrm{sim}}(f_0)/r_{\mathrm{sim}}(8.8\,\mathrm{nN})$. Time-scale factors are not defined for the lag-like zero baseline.}
\label{tab:benchmark}
\small
\begin{tabular}{lccccc}
\toprule
Reference & Lag & Phase 2 & $k_2$ & Mean velocity & Extent at 100\,min\\
          & (min) & (min) & (\%\,min$^{-1}$) & (\%\,min$^{-1}$) & (\%)\\
\midrule
Prakhya et al., Fig.~6A & 26 & 26--36 & 5.72 & 0.99 & 99.5\\
Prakhya et al., Fig.~2B & 26 & 26--34 & 1.81 & 0.81 & 82.2\\
\end{tabular}

\vspace{6pt}
\setlength{\tabcolsep}{4.5pt}
\begin{tabular}{cccccccc}
\toprule
$f_0$ & Post- & $r_{\mathrm{sim}}$ & $p$ & $t_{\mathrm{eq}}$ (min) & $\lambda_t$ & $\lambda_r$ & $r_{\mathrm{rel}}$\\
(nN) & lag & (\%\,min$^{-1}$) & 6A\,/\,2B & 6A\,/\,2B & 6A\,/\,2B & 6A\,/\,2B & \\
\midrule
0   & 1/9 & 0.56 [0.31, 0.90] & 0.001\,/\,0.002 & -- & -- & -- & 0.05\\
2.2 & 9/9 & 7.9 [6.7, 9.0]    & 0.024\,/\,0.030 & 0.41\,/\,2.00 & 1.7\,/\,8.0  & 1.4\,/\,4.4 & 0.61\\
4.4 & 9/9 & 11.5 [9.9, 13.1]  & 0.037\,/\,0.044 & 0.62\,/\,2.77 & 2.5\,/\,11.1 & 2.0\,/\,6.4 & 0.89\\
8.8 & 9/9 & 12.8 [11.5, 14.0] & 0.046\,/\,0.056 & 0.78\,/\,3.27 & 3.1\,/\,13.1 & 2.2\,/\,7.1 & 1\\
\bottomrule
\end{tabular}
\end{table}

\subsection{Escape kinetics respond heterogeneously across realizations}

\begin{figure}[t]
  \centering
  \includegraphics[width=0.48\textwidth]{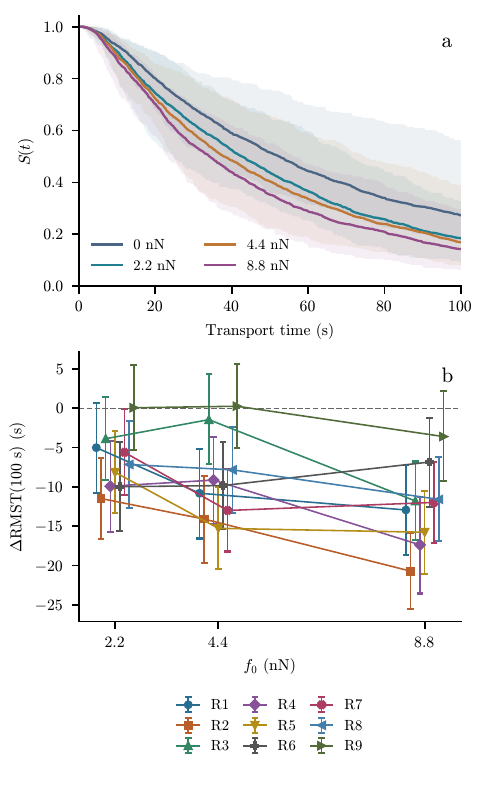}
  \caption{\textbf{Escape kinetics at $\bm{t_m=15}$\,s.} (a) Replica-mean survival $S(t)$ for the four baselines; shading spans the minimum--maximum range across the nine replicas and is not a confidence interval. (b) Within-replica difference in RMST restricted to 100\,s relative to the matched $f_0=0$ control, with approximate 95\% percentile-bootstrap intervals (10,000 resamples of paired tracer indices). Intervals are conditional on the network and starting sample and do not represent between-clot uncertainty.}
  \label{fig:escape}
\end{figure}

Despite the uniform geometric response, escape kinetics did not change uniformly (Fig.~\ref{fig:escape}). At $t_m=15$\,s, replica-mean survival at 100\,s decreased from 27.34\% at $f_0=0$ to 14.32\% at 8.8\,nN, and $\RMST(100)$ from 56.47 to 43.97\,s. All nine replicas had a lower RMST than their control at 8.8\,nN (reductions of 3.59--20.70\,s; sign test $p=0.0039$), and eight of the nine bootstrap intervals excluded zero. At 2.2 and 4.4\,nN, eight of nine replicas showed a reduction ($p=0.039$ each, uncorrected; $p=0.078$ after Holm correction across the three baselines). Replica 9 had small positive contrasts at both lower baselines, and replica 6 had a greater RMST at 8.8 than at 4.4\,nN while remaining below its control. The aggregate ordering of means therefore does not represent a monotonic dose--response shared by all realizations.

Accessible volume alone did not predict residence. At $t_m=10$\,s and $f_0=4.4$\,nN, $\phiacc$ increased by 3.21 percentage points in replica 1 and 3.02 points in replica 2, but the corresponding RMST changes were $+1.43$ and $-14.04$\,s. The zero-baseline state of replica 4 at 10\,s retained 60.55\% of tracers at 100\,s ($\RMST=78.36$\,s); this realization was retained in all summaries. For reference, the mean first-passage time from the midplane of an obstacle-free slab, $L_z^2/(8D)=7.8$\,s \citep{redner2001}, is 5.6--7.3 times smaller than the observed replica-mean RMST values, which are themselves lower bounds on the unrestricted mean exit time.

\subsection{Further mechanical evolution produces small and mixed changes}

\begin{figure}[t]
  \centering
  \includegraphics[width=0.46\textwidth]{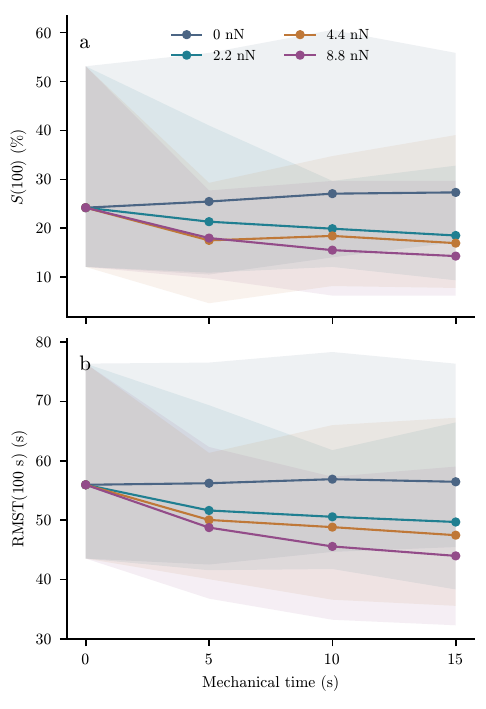}
  \caption{\textbf{Transport readouts versus mechanical time.} (a) Survival at 100\,s and (b) RMST restricted to 100\,s for networks frozen at 0, 5, 10 and 15\,s. Points are means over the same nine replicas; bands span the minimum--maximum range, not confidence intervals. The initial state is shared across baselines.}
  \label{fig:mechtime}
\end{figure}

Advancing the frozen state from 10 to 15\,s changed mean RMST by $-0.44$, $-0.87$, $-1.36$ and $-1.59$\,s at baselines 0, 2.2, 4.4 and 8.8\,nN, respectively (Fig.~\ref{fig:mechtime}). At the highest baseline, mean $S(100)$ decreased from 15.54\% to 14.32\% while mean $\Cint$ increased from 3.586\% to 4.602\%. These aggregate shifts were not consistent within replicas: of the 27 active-baseline 15-minus-10\,s comparisons, 14 had negative RMST point differences and five had intervals entirely below zero, with point differences spanning $-9.08$ to $+4.68$\,s. Because each comparison involves newly released populations in two frozen geometries, these values describe changes between states rather than trajectories followed through a moving network.

\subsection{Retention occurs predominantly in connected space}

To separate loss of exit routes from slow motion within connected space, every starting centre was classified independently of the observed trajectories using conservative inner and outer cell graphs (Section~\ref{sec:survival}). Across the 29,952 starts, 20,052 were certified as connected to at least one absorbing face, six were certified as disconnected and 9,894 (33\%) remained unresolved. Of the 6,100 histories still present at 100\,s, 3,380 started at certified connected centres, six at certified disconnected centres and 2,714 at unresolved centres. Thus, 3,380 of the 3,386 retained tracers with a classified start had an exit route. In replica 1 at $f_0=2.2$\,nN and $t_m=10$\,s, for example, 255 of 256 starts were certified connected and all 64 retained tracers belonged to that group, so their retention cannot be attributed to an isolated compartment.

Isolation nevertheless occurs. Replica 4 contained certified disconnected starts at 4.4 and 8.8\,nN at 10\,s and at 2.2 and 4.4\,nN at 15\,s; none of these tracers escaped within 100\,s, which provides a non-trivial consistency check of the classification. These six classifications involve two distinct starting coordinates reused across geometries rather than six distinct isolated pores. Conversely, replica 1 at 4.4 and 8.8\,nN and 10\,s left all 256 starts unresolved despite 184 and 205 observed escapes, illustrating that an unresolved start is not evidence of isolation.

\section{Discussion}

The main finding is a separation between geometric accommodation and escape kinetics. Increasing platelet traction produced more projected contraction and more accessible volume in every replica and at every mechanical time, but residence time responded heterogeneously, with a reduction shared by all replicas only at the highest baseline. Larger accessible volume can therefore coexist with a slowly escaping central population, and nearly all retained tracers with a classified start had a route to an exit. In this model, retention over 100\,s is dominated by the time required to traverse connected but tortuous space rather than by trapping in isolated pores, although isolated pores do occur. This is consistent with first-passage theory, in which mean exit times are controlled by the global geometry of the accessible domain and by narrow passages rather than by porosity alone \citep{benichou2014,holcman2014}.

These results have implications for how clot contraction is interpreted. Projected-area contraction is the standard experimental readout \citep{tutwiler2016,prakhya2022}, and greater contraction is often equated with a denser, less permeable clot \citep{wufsus2013}. Our simulations suggest that, at the early, low-strain stage modelled here, platelet traction concentrates fibrin around platelets and opens space between them, so that the clot becomes locally more accessible to micrometre-scale particles while its escape kinetics vary strongly between realizations of the same mechanical protocol. Particle retention by contracting clots, for example of microparticles \citep{zubairova2015} or therapeutic carriers, may therefore depend more on network connectivity and channel structure than on the extent of contraction itself.

Benchmarking against experimental clot-contraction kinetics supports the mechanical basis of these results. Without fitting any parameter to contraction data, the model separates traction-competent from lag-like states by the experimental lag criterion, produces contraction rates within one order of magnitude of experimental phase-2 rates and brackets the rate reduction produced by myosin inhibition \citep{prakhya2022}. The benchmark also defines the scope of the transport conclusions: mapped by strain, the analysed states correspond to the first 2--6\% of experimental contraction, at the onset of the contraction phase. Because a common rescaling of the drag coefficients changes only the time axis, the transport conclusions, which depend on the configurations reached, are unaffected by the mismatch in absolute rate; the benchmark constrains the stage of contraction, not the clock. Two discrepancies remain. Simulated contraction starts at its maximal rate and decelerates, whereas experimental contraction accelerates after the lag ($\lambda_t>\lambda_r$), consistent with all platelets being traction-competent at $t_m=0$ in the model but activated and recruited progressively in experiments \citep{tutwiler2016,prakhya2022}. Reproducing this shape would require an explicit motor time scale, such as a filopodial force--velocity relation, rather than a rescaling of the drag coefficients. In addition, the simulated platelet density ($\sim2\times10^{7}\,\mu\mathrm{L}^{-1}$) exceeds that of the reference experiment by about 400-fold, so the model represents a platelet-rich microdomain and the rate agreement should be read as order-of-magnitude.

Several limitations define the scope of these conclusions. First, the domain is assembled from independently evolved two-dimensional layers with an imposed spacing of 100\,nm, equal to the nominal fibre diameter; it is not a mechanically coupled three-dimensional reconstruction, and the 5-$\mu$m thickness fixes the axial exit distance. Second, mechanics were simulated for 15\,s, which by strain corresponds to the onset of the contraction phase; later compaction, which reaches much larger volumetric strains over minutes \citep{litvinov2023,tutwiler2016}, was not modelled. Third, tracers do not bind, interact or experience flow, and the diffusion coefficient is an assumed effective value; because the geometry is frozen and the dynamics are drift-free, survival curves for another diffusivity follow by rescaling time (Section~\ref{sec:tracers}). Fourth, nine mechanical replicas reveal substantial between-realization heterogeneity, and the conditional bootstrap intervals do not capture between-network uncertainty; replica-level exact tests support the geometric response and the highest-baseline transport response, but not a general dose--response. Finally, time-step equivalence was established in three tested states, but for survival in one of them only under a tolerance widened after inspection ($\pm2.5$ instead of $\pm2$\,pp), spatial convergence of the full-domain transport was not tested, and one third of starting centres remain unclassified by the connectivity certificate.

\subsection{Future work}
\label{sec:future}

\paragraph{Mechanical coupling and validation.} The most direct extensions are to couple the planes mechanically or use a fully three-dimensional network, to compare frozen-state transport with transport during mechanical evolution, which requires a rule for moving-boundary encounters, and to validate the structural predictions against microscopy.

\paragraph{Diffusion and escape in optical speckle landscapes.} A complementary, experimentally accessible analogue of the frozen-geometry problem is offered by laser speckle. When coherent light is transmitted through a thin film of clotted material, the multiply scattered field forms a random intensity pattern whose statistics are imprinted by the refractive-index heterogeneity of the fibrin--platelet network. Because dielectric microparticles experience optical gradient forces that scale with the local intensity, such a speckle field acts as a disordered optical potential, effectively a random optical lattice in which particles diffuse, are transiently trapped and escape. Speckle grains are not planar objects: they have a finite axial extent, and the three-dimensional structure of speckle fields, including clustered speckle generated with structured pupils, can be described theoretically, simulated and verified experimentally \citep{lencina2012}. Choosing the illumination and observation plane so that the axial grain length exceeds the tracer diameter would allow the landscape to be treated, to a first approximation, as a quasi-planar potential, directly analogous to a single layer of the present model, so that passage through the optical lattice reproduces in planar approximation passage through the clot. We therefore plan to (i) record speckle patterns transmitted through thin clot films prepared at different stages of contraction, (ii) convert the measured intensity into an effective potential landscape for Brownian tracers, and (iii) compute survival and restricted mean exit times in these landscapes with the same censoring-aware estimators used here. This would replace the hard-wall exclusion of the present model with a soft potential whose depth is tuned by laser power, bridging purely geometric accessibility and energetic trapping.

\paragraph{Platelet aggregation in neonatal clots.} A second direction concerns the developmental specificity of clot formation. Platelet aggregation and clot formation in the neonatal period have received far less attention than in adults, and models tailored to this stage of development are lacking, in part because only small blood volumes can be obtained from newborns. Optical techniques that work with microlitre samples are therefore attractive: dual time-resolved membrane fluctuation spectroscopy has already been used to characterize membrane flickering of fetal (umbilical-cord) erythrocytes under gestational diabetes \citep{bravo2023}, demonstrating that the mechanics of neonatal blood cells are accessible to optical probes. Adapting the present framework to neonatal conditions would require re-parameterizing the filopodial traction law (the baseline $f_0$ and the stiffness-dependent coefficients of Eq.~\ref{eq:traction}), platelet density and fibre properties, and representing platelet aggregation explicitly rather than as a fixed platelet population. Testing the resulting transport predictions requires following microparticles at high temporal resolution in small sample volumes. Event-based (neuromorphic) cameras combined with optical modulation of the image plane increase the temporal resolution of Brownian microparticle tracking by up to about 400-fold while preserving the stochastic statistics of the motion \citep{huenchual2025}, which would allow escape-time distributions measured in neonatal and adult clot films to be compared directly with simulated $S(t)$ and RMST.

Within its stated scope, the present study provides a reproducible framework that links platelet-driven mechanics to finite-size accessibility, certified connectivity and censoring-aware escape statistics, and shows that these three descriptors must be distinguished when inferring transport from contraction.

\section{Methods}
\label{sec:methods}

\subsection{Overdamped network and platelet dynamics}

Fibrin nodes and platelet centres were advanced with an overdamped Langevin update. For object $i$ with position $\mathbf{q}_i$, drag coefficient $\zeta_i$, total force $\mathbf{F}_i$ and diffusion coefficient $D_i=k_BT/\zeta_i$,
\begin{equation}
\mathbf{q}_i(t+\delta t)=\mathbf{q}_i(t)+\frac{\delta t}{\zeta_i}\,\mathbf{F}_i(t)+\sqrt{2D_i\,\delta t}\;\bm{\xi}_i,
\label{eq:langevin}
\end{equation}
where $\bm{\xi}_i\sim\mathcal{N}(\mathbf{0},\mathbf{I}_2)$ is sampled independently for every object and coordinate. Quantities are expressed in $\mu$m, nN and s. The effective drag coefficients were $\zeta_n=3.769911$\,nN\,s\,$\mu$m$^{-1}$ for fibrin nodes and $\zeta_p=85.2$\,nN\,s\,$\mu$m$^{-1}$ for platelets. The nodal force comprises backbone and cohesive worm-like-chain (WLC) forces, rest-angle bending, platelet--fibrin contact and filopodial traction; the platelet force comprises platelet--platelet interaction and the reactions to contact and traction. After each stochastic update the common translational displacement of all free objects was removed to suppress whole-network drift. Forces were recomputed in adaptive internal substeps,
\begin{equation}
\delta t=\min\!\left(\Delta t_{\mathrm{rem}},\,\frac{\delta x_{\mathrm{target}}}{v_{\max}}\right),\qquad v_{\max}=\max_i\frac{\|\mathbf{F}_i\|}{\zeta_i},\qquad \delta x_{\mathrm{target}}=0.01\um,
\end{equation}
where $\Delta t_{\mathrm{rem}}$ is the remainder of the 1-ms outer step. A proposed displacement larger than $1\um$ terminated the simulation with an error.

\subsection{Fibrin discretization and WLC mechanics}

Each initial fibrin edge of length $L$ was subdivided once into $S_{\mathrm{sub}}=\max(2,\lceil L/l_{\mathrm{seg}}\rceil)$ segments with $l_{\mathrm{seg}}=0.30\um$. Each segment $e$ stores its initial length $L_{0,e}$, and its state is described by
\begin{equation}
\gamma_e=\frac{L_e-L_{0,e}}{L_{0,e}},\qquad \hat\gamma_e=\min(\gamma_e,\gamma_{\max}),\qquad u_e=\frac{\hat\gamma_e}{C},
\end{equation}
where $C$ is a dimensionless extension parameter and $0<\gamma_{\max}<C$. Stretching follows the Marko--Siggia interpolation \citep{marko1995,michael2023} for $N$ chains in parallel with persistence length $P$,
\begin{equation}
F_{\mathrm{WLC},e}=\frac{Nk_BT}{P}\left[\frac{1}{4(1-u_e)^2}-\frac14+u_e\right],
\end{equation}
which vanishes at the reference length and gives a signed compressive response for $\gamma_e<0$. Its tangent,
\begin{equation}
k_{\gamma,e}=\left.\frac{\partial F_{\mathrm{WLC}}}{\partial\gamma}\right|_{\hat\gamma_e}=\frac{Nk_BT}{P}\left[\frac{C^2}{2(C-\hat\gamma_e)^3}+\frac1C\right],
\end{equation}
evaluated at the mean signed strain of the backbone segments incident on node $j$, $k_j=k_\gamma(\min(\bar\gamma_j,\gamma_{\max}))$, is the local stiffness signal for filopodial traction. Parameters were $T=300$\,K, $N=50$, $P=1.0740335\times10^{-3}\um$, $C=2.3512436$ and $\gamma_{\max}=2.0$ (Table~\ref{tab:S1}).

\subsection{Bending, contact, traction and cohesion}

For a node triplet $(i,j,k)$ with $\mathbf{u}=\mathbf{x}_j-\mathbf{x}_i$ and $\mathbf{v}=\mathbf{x}_k-\mathbf{x}_j$, the signed turning angle is $\varphi=\operatorname{atan2}\big((\mathbf{u}\times\mathbf{v})_z,\mathbf{u}\cdot\mathbf{v}\big)$, and its deviation from the stored angle $\varphi_0$, wrapped to $(-\pi,\pi]$, defines the bending energy
\begin{equation}
E^{\mathrm{bend}}_{ijk}=\frac12\,\frac{EI}{L_{0,ijk}}\,(\Delta\varphi)^2,\qquad L_{0,ijk}=\frac{L_{0,ij}+L_{0,jk}}{2},
\end{equation}
with $EI=2.2625934\times10^{-3}$\,nN\,$\mu$m$^2$. Triplets were assigned along fibre interiors and between angularly adjacent segments at junctions.

Platelets are disks of radius $r_p=1.13\um$; fibres have radius $r_f=0.05\um$. With $r_{\mathrm{in}}=f_rr_p+r_f$ ($f_r=0.9$) and $r_{\mathrm{out}}=r_p+r_f$, the force of platelet $m$ on node $j$ at distance $d_{mj}$ along unit vector $\hat{\mathbf{r}}_{mj}$ is repulsive, $+F_{PF}\hat{\mathbf{r}}_{mj}$, for $d_{mj}<r_{\mathrm{in}}$; adhesive, $-F_{PF}\hat{\mathbf{r}}_{mj}$, for $r_{\mathrm{in}}\le d_{mj}\le r_{\mathrm{out}}$; and zero otherwise, with $F_{PF}=4$\,nN and an equal and opposite reaction on the platelet. Filopodia act in the annulus $r_{\mathrm{out}}\le d_{mj}\le r_{\max}=2.18\um$. Each platelet carries a lognormally distributed number of arms (log-space $\mu=1.85$, $\sigma=0.27$, rounded and clipped to 2--13). A free arm attaches to the nearest unoccupied node in the annulus and is released and reattached when its node leaves the annulus; no node can be claimed by two arms. Each attached arm applies
\begin{equation}
F_{\mathrm{fil}}(k_j)=\max\!\left[0,\;\frac12\left(f_0+\frac{a\,k_j}{b\,k_j+c}\right)\right],
\label{eq:traction}
\end{equation}
with $a=61$\,nN, $b=3.582$ and $c=131.8$\,nN, and an equal and opposite reaction on the platelet. Setting $f_0=0$ removes only the baseline term.

Platelet pairs interact through a truncated 12--6 Lennard-Jones force with core diameter $d_c=2f_rr_p$, $\sigma=d_c/2^{1/6}$, cutoff $r_c=2r_p$ and $\epsilon=4$\,nN\,$\mu$m, repulsive for $r<d_c$ and attractive for $d_c<r<r_c$. Every ten outer steps, cohesive links were created between nodes of distinct fibres closer than $r_{\mathrm{coh}}=0.10\um$, excluding existing pairs, with at most four new links per node per search and at most four cohesive neighbours per node. A link created at time $t_b$ takes its current length as reference length, so that its force is zero at creation, and thereafter follows the same WLC law as a backbone segment; links are permanent.

\subsection{Mechanical cohort}

Each layer contains 45 platelets in a $47.6\times47.6\um^2$ domain, with degree-3 and degree-4 junction densities of 0.020 and 0.00379\,$\mu$m$^{-3}$ converted to planar node counts using an initialization thickness of $1\um$. Nine replicas were generated, each consisting of 50 independently seeded layers; replicas 1 and 2 were assembled from 20-layer stacks extended by 30 layers under the same protocol, and replicas 3--9 were generated directly with 50 layers. Each replica was evolved at the four baselines from a shared initial state, with output every 0.1\,s; evolution to 15\,s continued from saved 10-s states, retaining cohesive links, reference lengths and the stochastic state. Contraction is always referenced to the 0-s configuration. All inputs are listed in Table~\ref{tab:S1}; none were fitted to the transport results.

\subsection{Time scale of the mechanical model}
\label{sec:timescale}

Equation~\eqref{eq:langevin} contains no time scale other than that set by the drag coefficients together with forces and stiffnesses; it includes no platelet activation time, actomyosin force--velocity relation or filopodial binding kinetics. Multiplying all $\zeta_i$ by a common factor $\alpha$ and rescaling time as $t\to\alpha t$ leaves Eq.~\eqref{eq:langevin} unchanged, including the noise term because $D_i=k_BT/\zeta_i$, so the network traverses the same sequence of configurations $\alpha$ times more slowly; the equivalence is exact apart from events scheduled at fixed model-time intervals, such as the cohesion search. The drag coefficients exceed Stokes drag in plasma ($\eta=1.2$\,mPa\,s) by $\sim1.6\times10^{6}$ for a 0.3-$\mu$m fibre segment and $\sim3.3\times10^{6}$ for a platelet. They are therefore effective parameters that absorb the slow kinetics of platelet activation and actomyosin contraction rather than hydrodynamic friction, and thermal displacements are negligible ($\sim6$\,nm rms per coordinate in 15\,s for a fibrin node). Combined with the model forces, $\zeta_n$ sets two intrinsic time scales: elastic relaxation of a WLC segment, $\tau_{\mathrm{WLC}}=\zeta_nl_{\mathrm{seg}}/k_\gamma(0)\approx9$\,s with $k_\gamma(0)=3Nk_BT/(2PC)=0.123$\,nN, and filopodial capture across the $\sim1$-$\mu$m reach, $\tau_{\mathrm{fil}}\approx\zeta_n\times1\um/F_{\mathrm{fil}}=0.85$, 1.7 and 3.3\,s at $f_0=8.8$, 4.4 and 2.2\,nN. At $f_0=0$ the stiffness-dependent term of Eq.~\eqref{eq:traction} contributes only $\approx0.03$\,nN per arm at rest, giving $\tau_{\mathrm{fil}}\approx130$\,s. Mechanical time $t_m$ thus labels the progression of contraction rather than calibrated physical time. Frozen-state transport depends only on the configurations reached, and the factors $\lambda_r$ and $\lambda_t$ defined below quantify the rescaling required to match experimental kinetics.

\subsection{Comparison with experimental contraction kinetics}
\label{sec:benchmark}

Simulated contraction was compared with the time-lapse projected-area kinetics of \citet{prakhya2022} (washed mouse platelets, $5\times10^{7}$\,mL$^{-1}$; fibrinogen 250\,$\mu$g\,mL$^{-1}$; thrombin 20\,mU\,mL$^{-1}$; $\sim25\,^\circ$C), whose observable, clot area as a percentage of the initial area, matches the normalization of $\Cint$. Their two standard-condition control curves (Figs.~2B and 6A of \citealp{prakhya2022}) were digitized from the published figures, resampled to 1-min frames and analysed with their protocol: the lag ends at the first frame-to-frame area decrease exceeding 1\%, here required to persist for three consecutive frames to exclude an initial settling drop and digitization noise, and the contraction phase (phase 2), delimited by the extrema of the derivative smoothed over ten neighbours on each side, is fitted linearly to give $k_2$ \citep{tutwiler2016,prakhya2022}. Because model time is set by effective drag coefficients, simulations were mapped onto experiments by strain and rate rather than by identifying simulated with experimental seconds. For each replica we computed the fraction of the experimental extent $C_{\mathrm{ext}}$ reached at 15\,s, $p=\Cint(15\,\mathrm{s})/C_{\mathrm{ext}}$; the post-lag time $t_{\mathrm{eq}}$ at which the experiment reaches the same contraction; and two time-scale factors, $\lambda_t=t_{\mathrm{eq}}/15$\,s and $\lambda_r=r_{\mathrm{sim}}/k_2$, where $r_{\mathrm{sim}}$ is the linear contraction rate between 5 and 15\,s. Agreement between $\lambda_t$ and $\lambda_r$ would indicate curves of the same shape. Intervals are percentile bootstrap intervals over the nine replicas.

\subsection{Solid assembly and accessible domain}

For mechanical time $t_m$, let $\mathcal{A}_k(t_m)$ be the padded analysed support of layer $k$. The common lateral region and the transport domain are
\begin{equation}
\mathcal{A}_\cap(t_m)=\bigcap_{k=1}^{N_z}\mathcal{A}_k(t_m),\qquad \mathcal{A}=\mathcal{A}_\cap\times[0,L_z].
\end{equation}
Backbone and cohesive segments are represented as planar capsules of radius $r_f$ and platelets as disks of radius $r_p$, extruded over their layer slab; their union is the solid $\mathcal{S}$. The lateral boundary retains the native $256\times256$ raster (spacing $0.1859375\um$) and reflects tracer centres; solid exclusion uses the continuous centre-to-solid distance, so that the accessible set is
\begin{equation}
\Omega_{\mathrm{acc}}=\{\mathbf{r}\in\mathcal{A}:\ \mathrm{dist}(\mathbf{r},\mathcal{S})\ge a_{\mathrm{tr}}\}.
\end{equation}
Contraction was quantified as $\Cint(t_m)=1-|\mathcal{A}_\cap(t_m)|/|\mathcal{A}_\cap(0)|$. These quantities describe projected analysed regions, not volumetric strain. The accessible fraction $\phiacc=|\Omega_{\mathrm{acc}}|/|\mathcal{A}|$ and volume $V_{\mathrm{acc}}=\phiacc|\mathcal{A}|$ were estimated from 131,072 uniform proposals per replica shared across its states. Accessibility includes isolated components and is not an escape probability.

\subsection{Tracer transport}
\label{sec:tracers}

In each geometry, 256 noninteracting spherical tracers ($a_{\mathrm{tr}}=0.25\um$, $D=0.4\um^2\,\mathrm{s}^{-1}$) were initialized uniformly in a central cylinder of radius $1.19\um$ with $2\le z\le3\um$, conditioned on solid clearance but not on connectivity (Fig.~\ref{fig:S1}). The diameter matches the size scale of circulating microparticles \citep{zubairova2015}; $D$ is an assumed effective value. Free increments $\Delta\mathbf{R}=\sqrt{2D\Delta t}\,\bm{\eta}$, $\bm{\eta}\sim\mathcal{N}(\mathbf{0},\mathbf{I}_3)$, with $\Delta t=0.015625$\,ms, were checked for the first contact with the radius-expanded solid or the lateral boundary, and the remaining displacement was reflected about the contact normal,
\begin{equation}
\Delta\mathbf{R}'_{\mathrm{rem}}=\Delta\mathbf{R}_{\mathrm{rem}}-2(\Delta\mathbf{R}_{\mathrm{rem}}\cdot\mathbf{n})\,\mathbf{n},
\end{equation}
with further contacts resolved within the same increment. This is a drift-free geometric reflection scheme rather than the force-dependent algorithm of \citet{behringer2012}. Escape was recorded at the end of the step in which a centre first reached $z=0$ or $z=L_z$. Because the geometry is frozen and the dynamics are drift-free, results for another diffusivity $D'$ follow, up to discretization error, from $S_{D'}(t)=S_D(tD'/D)$ and $\RMST_{D'}(T)=(D/D')\,\RMST_D(TD'/D)$.

\subsection{Survival, restricted mean exit time and connectivity}
\label{sec:survival}

With first-passage time $\tau$ to either axial face, survival is $S(t)=\Pr(\tau>t)$ \citep{redner2001}. All histories end at escape or at $T_c=100$\,s, so that
\begin{equation}
\hat S(t)=1-\frac{N_{\mathrm{esc}}(t)}{N_0},\qquad \widehat{\RMST}(T_c)=\frac{1}{N_0}\sum_i\min(\tau_i,T_c)=\int_0^{T_c}\hat S(t)\,dt.
\end{equation}
Censored histories contribute $T_c$; the RMST \citep{royston2013} is used without a hazard model, and finite-time survival is not interpreted as permanent capture. For a fully resolved connected fraction $F_{\mathrm{conn}}$ of starting centres,
\begin{equation}
S(t)=1-F_{\mathrm{conn}}+F_{\mathrm{conn}}S_{\mathrm{conn}}(t),\qquad \RMST(T_c)=(1-F_{\mathrm{conn}})\,T_c+F_{\mathrm{conn}}\RMST_{\mathrm{conn}}(T_c),
\label{eq:decomp}
\end{equation}
which separates loss of exit routes from slow motion in connected space. Connectivity of each start to at least one absorbing face was assessed independently of the trajectories with conservative inner and outer cell graphs on a $512\times512\times100$ auxiliary grid and checked continuous bridges (Supplementary Note~\ref{note:current}). With $n_+$ connected, $n_-$ disconnected and $n_?$ unresolved starts, $n_+/N_0\le\hat F_{\mathrm{conn}}\le1-n_-/N_0$; an uncertified path does not establish isolation.

\subsection{Paired comparisons and statistics}

Each active-baseline state was compared with the zero-baseline control of the same replica $r$ at the same mechanical time, $\Delta Y_r(f_0,t_m)=Y_r(f_0,t_m)-Y_r(0,t_m)$. Initial proposal pools and per-particle Gaussian seeds were reused within each replica; because validity filtering depends on geometry, control and active states shared 26--225 of 256 physical starting positions. Within-replica transport contrasts used 10,000 percentile-bootstrap resamples of paired particle indices \citep{efron1979}; these 95\% intervals are approximate, marginal and conditional on the network and starting sample. Inference across realizations treated the nine replicas as independent units and used the exact two-sided sign test on the signs of $\Delta Y_r$ (minimum attainable $p=2^{-8}\approx0.0039$), with Holm correction across conditions where stated. No inference treats tracers or layers as independent clots.

\subsection{Numerical verification}

The reflecting implementation passed 26,510 geometric and propagation tests, including analytic first contacts, independent segment--box comparisons, planar image solutions and multiple-reflection controls; a 4096-tracer straight-channel benchmark reduced the relative RMST error from 5.28\% at 0.5\,ms to 1.17\% at 0.015625\,ms (Supplementary Note~\ref{note:current}). Time-step equivalence was assessed with paired histories that share accepted starting positions and nested Brownian increments across step sizes, using margins of $\pm2$ percentage points for survival at 100\,s and $\pm3\%$ for relative RMST, fixed in advance \citep{lakens2017}. An initial comparison on one geometry (256 paired histories, 0.03125 versus 0.015625\,ms) was inconclusive because of its small sample (survival difference $-1.953$\,pp, 97.5\% interval $[-12.238,\,8.408]$; RMST $-0.355\%$, $[-6.865,\,6.650]\%$). A fixed-allocation study then used 8192 paired histories in each of three frozen states (replica 1 at 0\,s; replica 1 at 2.2\,nN and 10\,s; replica 9 at 8.8\,nN and 15\,s) at 0.03125, 0.015625 and 0.0078125\,ms (73,728 trajectories; Table~\ref{tab:temporal-revised}). For the primary contrast, 0.015625 minus 0.0078125\,ms, relative RMST differences were 0.15--0.34\% with all intervals within $[-1.07,\,1.39]\%$, and survival differences were $-0.05$ to 0.81\,pp with intervals within $[-1.42,\,2.39]$\,pp. RMST met its margin in all three states; survival met the original $\pm2$\,pp margin in two states, and in the third (upper limit 2.39\,pp) only under a margin of $\pm2.5$\,pp adopted after inspection of the results. The secondary 0.03125-minus-0.015625\,ms contrasts met the original margins in all states. The 0.015625-ms step was therefore used throughout; this conclusion applies to the tested states and does not certify uniform equivalence of $S(t)$ over all 117 geometries. Earlier fixed-mask and local continuous-geometry studies, including analytical first-passage controls and spatial-refinement tests, are documented in Supplementary Notes~S1--S7. Two of the 29,952 histories contacted the artificial lateral boundary (replica 8 at 2.2\,nN and 10\,s, and at 8.8\,nN and 15\,s), so direct lateral reflections were rare.

\section*{Declarations}

\paragraph{Acknowledgements.} The authors acknowledge the FONDEF project IT24i0064 and the Departamento de F\'isica and the Facultad de Ciencias F\'isicas y Matem\'aticas of the Universidad de Concepci\'on for providing the infrastructure and GPU equipment that made it possible to begin the computationally demanding simulations reported here.

\paragraph{Data availability.} All data supporting the findings of this study are reported in the article and its Supplementary Information (Supplementary Notes~S1--S8, Tables~S1--S10 and Figs.~S1--S6), including model inputs, verification studies, connectivity certification and the provenance of the analysed cohort. The event histories of the 29,952 production and 73,728 temporal-verification trajectories are distributed with the code repository (see Code availability).

\paragraph{Code availability.} The simulation and analysis code (layered platelet--fibrin mechanics, continuous reflecting tracer transport and numerical verification) is available at \url{https://github.com/1250fp/platelet-fibrin-transport}.

\paragraph{Author contributions.} J.P.S.-V. conceived the study, supervised the work and wrote the manuscript. P.T.G. developed the model and simulation code, performed the simulations and analysis, and prepared the figures. Both authors reviewed the manuscript.

\paragraph{Competing interests.} The authors declare no competing interests.


\clearpage
\appendix
\renewcommand{\thefigure}{S\arabic{figure}}
\renewcommand{\thetable}{S\arabic{table}}
\setcounter{figure}{0}
\setcounter{table}{0}
\setcounter{equation}{0}
\setcounter{section}{0}
\renewcommand{\theequation}{S\arabic{equation}}
\renewcommand{\thesection}{S\arabic{section}}
\newcommand{\suppnote}[1]{\refstepcounter{section}\section*{Supplementary Note \thesection: #1}\addcontentsline{toc}{section}{Supplementary Note \thesection: #1}}

\section*{Supplementary Information}
\addcontentsline{toc}{section}{Supplementary Information}

\begin{figure}[h]
  \centering
  \includegraphics[width=0.45\textwidth]{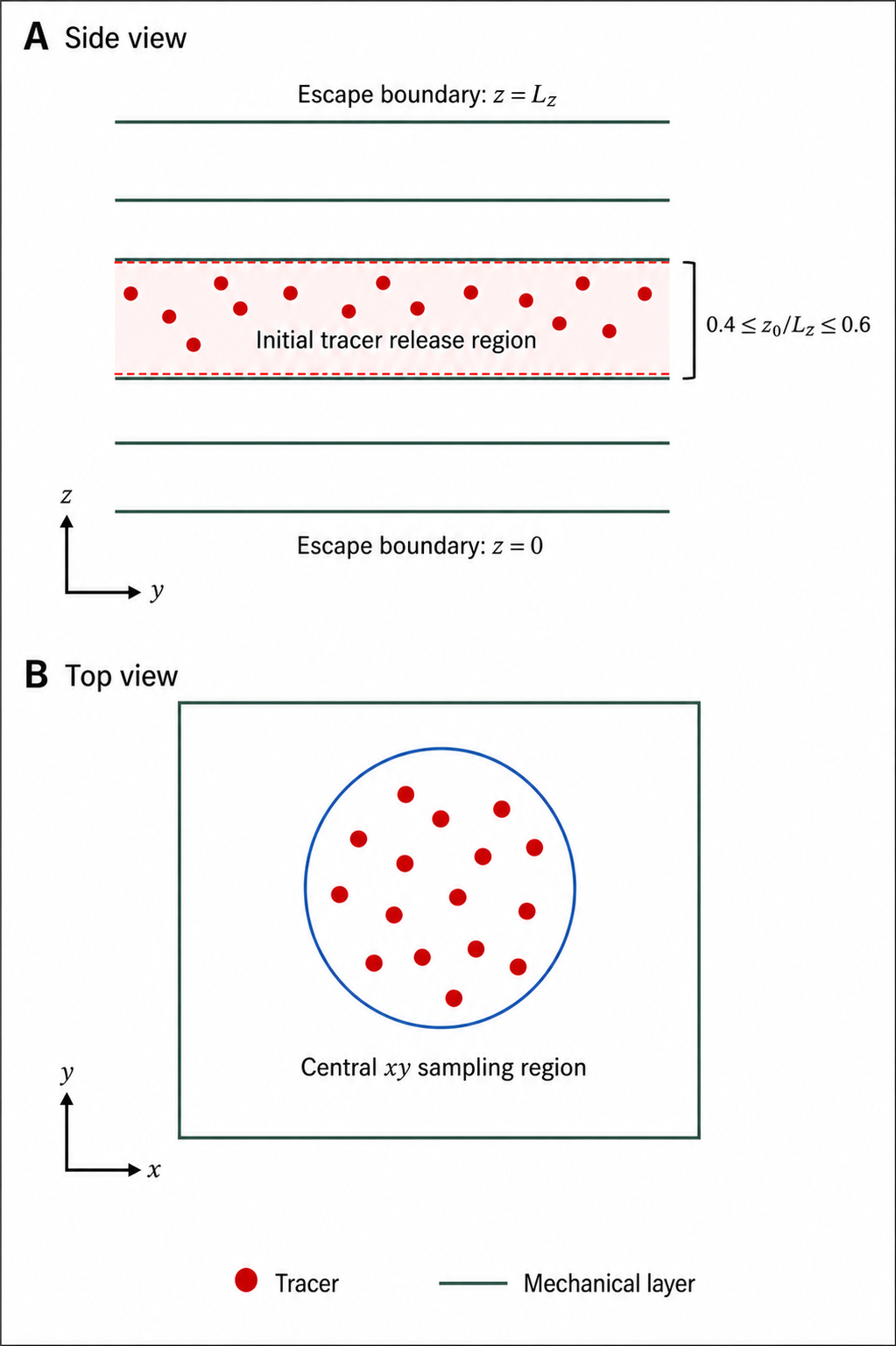}
  \caption{\textbf{Tracer release and boundary conditions.} (a) Side view: tracers are released in the band $0.4\le z_0/L_z\le0.6$ and escape when their centre first reaches $z=0$ or $z=L_z$. (b) Top view: starting centres are sampled in a central disk of radius $1.19\um$; the analysed lateral boundary reflects.}
  \label{fig:S1}
\end{figure}

This appendix contains the tracer release geometry (Fig.~\ref{fig:S1}), the complete list of model inputs and transport settings (Table~\ref{tab:S1}), the historical fixed-mask and local continuous-geometry verification studies (Supplementary Notes~S1--S7) and the reflection protocol, temporal-refinement study, connectivity certification and provenance of the main-text cohort (Supplementary Note~\ref{note:current}).

\begin{table}[h]
\centering
\caption{\textbf{Model inputs and transport protocol.} $N$, number of parallel WLC chains; $N_z$, number of layers.}
\label{tab:S1}
\small
\begin{tabular}{>{\raggedright\arraybackslash}p{0.33\textwidth}>{\raggedright\arraybackslash}p{0.6\textwidth}}
\toprule
Setting & Value or rule\\
\midrule
\multicolumn{2}{l}{\textit{Mechanical network}}\\
Planar domain; platelets per layer & $47.6\times47.6\um^2$; 45\\
Platelet and fibre radii & $r_p=1.13\um$; $r_f=0.05\um$\\
Junction densities (degree 3, 4) & 0.020 and 0.00379\,$\mu$m$^{-3}$; initialization thickness $1\um$\\
Target segment length & $l_{\mathrm{seg}}=0.30\um$\\
Temperature; parallel chains & $T=300$\,K; $N=50$\\
WLC parameters & $P=1.0740335\times10^{-3}\um$; $C=2.3512436$; $\gamma_{\max}=2$\\
Bending rigidity & $EI=2.2625934\times10^{-3}$\,nN\,$\mu$m$^2$\\
Drag (node; platelet) & 3.769911; 85.2\,nN\,s\,$\mu$m$^{-1}$\\
Cohesion & $0.10\um$; every 10 outer steps; valence 4\\
Platelet--fibrin force; pair energy & $F_{PF}=4$\,nN; $\epsilon=4$\,nN\,$\mu$m\\
Core fraction; filopodial reach & $f_r=0.9$; $r_{\max}=2.18\um$\\
Filopodial baseline & $f_0\in\{0,2.2,4.4,8.8\}$\,nN\\
Traction coefficients & $a=61$\,nN, $b=3.582$, $c=131.8$\,nN\\
Arms per platelet & lognormal, $\mu=1.85$, $\sigma=0.27$, clipped to 2--13\\
Integration & outer step 1\,ms; adaptive target $0.01\um$; output every 0.1\,s to 15\,s\\
Analysed support & padded convex support ($0.50\um$); $256\times256$ raster per layer\\
\midrule
\multicolumn{2}{l}{\textit{Frozen-state transport}}\\
Layers and thickness & $N_z=50$; $H=100$\,nm; $L_z=5.00\um$\\
Replicas and states & 9 per baseline; $t_m=0$, 5, 10, 15\,s\\
Tracer radius; diffusion coefficient & $a_{\mathrm{tr}}=0.25\um$; $D=0.40\um^2\,\mathrm{s}^{-1}$\\
Initial population & central cylinder, radius $1.19\um$, $2\le z\le3\um$; valid-centre conditioning\\
Histories & 256 per geometry; 117 geometries; 29,952 histories\\
Time step; horizon & $\Delta t=0.015625$\,ms; $T_c=100$\,s\\
Boundary conditions & reflection at solids and lateral boundary; absorption at both $z$ faces\\
Accessibility sampling & 131,072 uniform proposals per replica\\
Connectivity grid & $512\times512\times100$ cells; $h_{xy}=92.97$\,nm, $h_z=50$\,nm\\
\bottomrule
\end{tabular}
\end{table}

\clearpage
\suppnote{Scope, geometry, and implementation}
This supplement distinguishes implementation checks, analytical first-passage benchmarks, and discretization sensitivity. These questions are not equivalent to biological validation \citep{slater2021}. Supplementary Notes~S1--S7 preserve the historical fixed-mask and local continuous-geometry experiments with whole-proposal rejection. Supplementary Note~\ref{note:current} documents the subsequent full-common-domain reflection protocol, including cohesive solids, used for the homogeneous cohort of the main text. No simulation was rerun to compile this supplement.

The historical phase is closed provisionally, with an unresolved long-horizon escape-equivalence interval. Its conclusions remain conditional on the tested geometry and boundary rule and are not pooled with, or transferred automatically to, the current main-text cohort.

\subsection*{Two different geometric settings}
The fixed-mask temporal study uses 50 layers of thickness $H=70$ nm, total thickness $L=3.50~\mu$m, and a $47.6\times47.6~\mu$m field rasterized as $50\times256\times256$ cells. Its accessible-center mask is based on anisotropic EDT distances between raster sites.

The subsequent geometry and transport studies use a central $5.95\times5.95\times5.00~\mu$m region of the same saved 50-layer mechanical realization (base seed 4000, $f_0=2.2$ nN, mechanical time 10 s). Here $H=100$ nm is the assumed physical slab thickness, whereas $h_z=H/m$ is a numerical subdivision. The reference therefore contains 50 mechanical layers even when it contains 1600 numerical cells in $z$.

Each backbone segment is represented by a planar capsule, and each platelet by a planar disk, extruded across its own slab. Cohesive links are not added as new solid primitives in this geometry audit. This is a continuous interpretation of a layered model, not a reconstructed curved 3D fiber network or a set of spherical platelets. A halo of neighboring obstacles is retained for clearance queries; 2186 primitives enter the local detector.

The two size scenarios use $(a,D)=(0.10~\mu\mathrm{m},1.0~\mu\mathrm{m}^2/\mathrm{s})$ and $(0.25~\mu\mathrm{m},0.4~\mu\mathrm{m}^2/\mathrm{s})$. They change both steric size and diffusion. The final two refinements concern only the larger tracer.

\subsection*{Tracer update and boundary ordering}
All tracers are noninteracting and released inside the domain, not injected through $z=0$. Independent Gaussian coordinates generate each free proposal,
\begin{equation}
\mathbf R^*=\mathbf R+\sqrt{2D\Delta t}\,\boldsymbol{\eta},
\qquad \boldsymbol{\eta}\sim\mathcal N(\mathbf 0,I_3).
\end{equation}
The entire straight proposal is checked up to its first external boundary. A voxel supercover visits every touched cell, including edge/corner contacts. The continuous detector instead checks swept-sphere clearance against the extruded solid. A solid or lateral contact before, or tied with, an axial exit causes whole-proposal rejection and consumes the step. Otherwise the first free $z$ crossing absorbs the tracer. Escape time is recorded at the step endpoint, not as an exact Brownian first-contact time.

In the local test, $z=0$ and $z=5~\mu$m are absorbing for the tracer center, while the artificial ROI sides reject crossing proposals. Those sides can remove paths that would exist in the complete clot. Network mechanics remains frozen throughout the 100 s transport clock.

\subsection*{Implementation checks}
Checks include 16 transport-kernel tests of segment traversal, anisotropic cells, edges/corners, face ordering, obstacles before exit, final-step duration, and censoring. Independent continuous checks include 1200 segment-distance minimizations, analytic tangency cases, 4096 point-clearance comparisons, and 20000 detector comparisons before each recent refinement. The continuous distance search uses a bounding-volume hierarchy and conservative clearance bounds; it does not infer safety solely from free endpoints. These tests verify specified rules, not exact reflected Brownian motion.

\suppnote{Estimands, uncertainty, and equivalence}
Let $T_i$ be first exit through either axial face and let $T_c$ be the common observation horizon. With administrative right censoring,
\begin{align}
\widehat P_{\rm esc}(t)&=N^{-1}\sum_i\mathbf 1(T_i\le t),\\
\widehat S(t)&=1-\widehat P_{\rm esc}(t),\\
\widehat{\mathrm{RMST}}(T_c)&=N^{-1}\sum_i\min(T_i,T_c)
=\int_0^{T_c}\widehat S(t)\,dt .
\end{align}
Every censored history contributes $T_c$. The mean is not conditioned on escape, and positive $S(T_c)$ does not establish permanent capture \citep{redner2001,royston2013}. Exact integration of each empirical step-survival curve was checked against its reported RMST.

All differences are the coarser method minus the finer reference. Working margins were fixed at $\pm2$ percentage points (pp) for escape probability and $\pm3\%$ for relative RMST. Equivalence requires the complete 95\% confidence interval to fall inside the margin. A crossing interval is inconclusive; an interval entirely beyond a margin is outside. Joint equivalence requires \emph{both} indicators to pass at the specified horizon.

These tolerances are exploratory numerical choices, not universal physiological thresholds. A small point estimate or a nonsignificant difference from zero is insufficient to establish equivalence \citep{lakens2017}. The 95\% containment rule is more stringent than the 90\% interval convention used for standard TOST at the 5\% level; it is not presented here as that standard TOST procedure.

\subsection*{Sampling units}
The $H=70$ nm study uses paired differences of 16 batch means (1000 tracers per batch), with Student intervals on 15 degrees of freedom. Its seeds and starting sites are paired, but increments are not coupled step by step across resolutions.

The $H=100$ nm precision and nested-refinement studies use 8192 individually paired histories (32 batches of 256) on one fixed geometry. For a paired observable $Y$, the interval is
\begin{equation}
\overline{\Delta Y}\ \pm\ t_{0.975,N-1}
\frac{s_{\Delta Y}}{\sqrt N}.
\end{equation}
Relative RMST divides this interval by the observed mean of the finer method. A delta-method sensitivity includes uncertainty of that denominator; it changes none of the decisions in the final refinement. These intervals characterize tracer Monte Carlo uncertainty, not between-clot or between-mechanical-seed variability.

Wilson intervals describe pointwise survival. At zero observed discordance, a conservative exact bound avoids assigning zero uncertainty to an escape difference. Pointwise bands do not provide simultaneous coverage of the entire survival curve. Multiple horizons share histories; successive refinement stages are exploratory rather than a familywise-controlled sequence.

\subsection*{Reuse without double counting}
The original local pilot had 512 histories per size and method. The later precision study used 8192 new initial centers per size; its new-only analysis is primary, and pooled pilot-plus-new results are separate sensitivity analyses. The 0.03125 and 0.015625 ms stages each reuse those same 8192 centers and earlier histories. They add methods, not independent mechanical replicates or additional pairs to the primary sample.

\clearpage
\suppnote{Fixed-mask temporal verification at \texorpdfstring{$H=70$\,nm}{H = 70 nm}}
The primary comparison is $\Delta t=0.125$ versus $0.0625$ ms at $T_c=10,20,50,100$ s, for both tracer radii. Each radius and step uses 16000 trajectories, giving 64000 histories in total. Initial accessible voxel centers lie in the axial band $0.4L\le z<0.6L$ and a central in-plane disk. The same $50\times256\times256$ EDT mask is held fixed.

\begin{table}[!htbp]
\centering\small
\setlength{\tabcolsep}{4pt}
\caption{Fixed-mask temporal comparison: 0.125 minus 0.0625 ms. Both escape and RMST meet the working margins for all listed conditions. Intervals are paired 95\% confidence intervals across 16 Monte Carlo batches.}\label{tab:h70}
\begin{tabular}{@{}cccc@{}}
\toprule
$d$ ($\mu$m) & $T_c$ (s) & $\Delta P$ [CI] (pp) & $\Delta\mathrm{RMST}$ [CI] (\%) \\ \midrule
0.20 & 10 & $-0.281\ [-0.917,0.355]$ & $0.278\ [-0.763,1.320]$ \\
0.20 & 20 & $-0.031\ [-0.344,0.281]$ & $0.495\ [-0.863,1.852]$ \\
0.20 & 50 & $-0.006\ [-0.030,0.017]$ & $0.497\ [-0.938,1.931]$ \\
0.20 & 100 & $0.000\ [-0.031,0.031]$ & $0.496\ [-0.946,1.939]$ \\
0.50 & 10 & $-0.100\ [-0.951,0.751]$ & $0.246\ [-0.240,0.733]$ \\
0.50 & 20 & $-0.531\ [-1.504,0.441]$ & $0.477\ [-0.178,1.132]$ \\
0.50 & 50 & $-0.906\ [-1.771,-0.041]$ & $1.264\ [0.360,2.168]$ \\
0.50 & 100 & $-0.169\ [-0.713,0.376]$ & $1.618\ [0.323,2.913]$ \\
\bottomrule
\end{tabular}
\end{table}

All listed paired intervals meet the selected margins. The tightest case is the large-tracer RMST at 100 s: $1.618\%$ with interval $[0.323,2.913]\%$. At the fine step, the small tracer has no censored histories at 100 s, while the large tracer retains 1324/16000 histories, with RMST $35.5467$ s. Of the large-tracer starts, 198/16000 lie in components connected to neither axial face; none of those escaped.

The combined sample was fixed before completing its extension, guided by a pilot. A sensitivity analysis restricted to additional batches leaves some large-tracer comparisons inconclusive. It is not an independent confirmation of all the combined-sample decisions.

\begin{figure}[!htbp]
\centering\includegraphics[width=\textwidth]{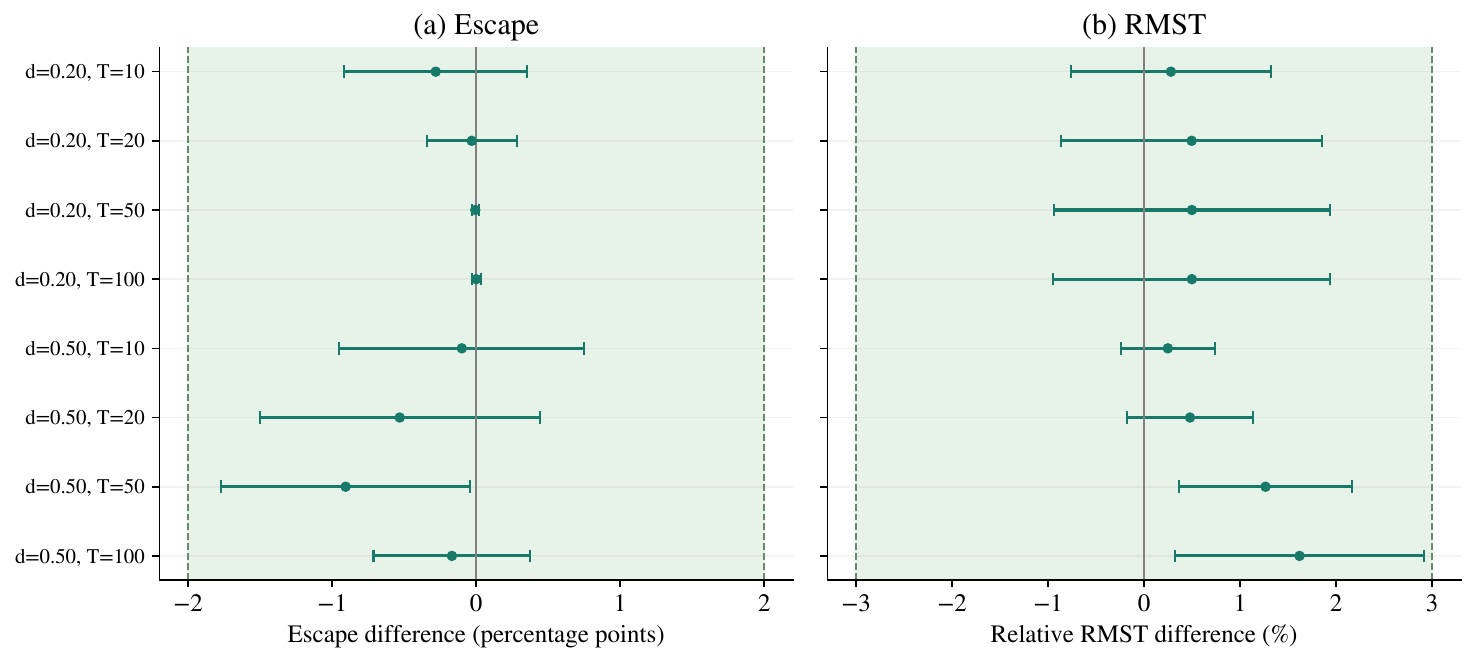}
\caption{Fixed-mask temporal equivalence. Labels give diameter $d$ in $\mu$m and horizon $T$ in seconds. Bars are paired 95\% intervals across batches; shaded regions are the working margins.}
\end{figure}

This result supports a finite-step comparison on the specified EDT mask. It does not validate the 1 ms endpoint pilot, exact sphere--surface clearance, the $H=100$ nm geometry, other mechanical realizations, or moving obstacles. A five-step exploratory sweep at 10 s showed larger-step sensitivity; its 4000 histories per condition are not added again to the 16000-history primary cohort.

\clearpage
\suppnote{Analytical control geometries}
The backward equation $D u''=-1$ provides mean-exit references in separable domains \citep{redner2001}. For an interval with absorbing endpoints,
\begin{equation}
u(z)=\frac{z(L-z)}{2D}.
\end{equation}
For a reflecting wall at accessible coordinate $a$ and an absorbing endpoint $L$, with $\ell=L-a$,
\begin{equation}
u(z)=\frac{\ell^2-(z-a)^2}{2D},\qquad u'(a)=0,\quad u(L)=0.
\end{equation}
A straight channel with ideal reflecting lateral walls has the same axial first-passage law as the free interval. Whole-proposal rejection can instead suppress an axial displacement when its transverse part touches a wall, producing finite-step delay.

The absorbing-interval survival expansion is
\begin{equation}
S(t\mid z)=\frac4\pi\sum_{j=0}^\infty
\frac{\sin[(2j+1)\pi z/L]}{2j+1}
\exp\!\left[-D\left(\frac{(2j+1)\pi}{L}\right)^2t\right].
\end{equation}
For the mixed reflecting/absorbing case, with $x=z-a$ and $q_j=(j+1/2)\pi/\ell$,
\begin{equation}
S(t\mid x)=\sum_{j=0}^\infty
\frac{2(-1)^j}{\ell q_j}\cos(q_jx)e^{-Dq_j^2t}.
\end{equation}
The references are averaged over the actual sampled starting positions and integrated to the same $T_c$ as the simulation. They do not use the numerical collision detector. For the voxel controls, 512 and 1024 modes agreed in RMST to better than $10^{-8}$ s.

\subsection*{Voxel controls}
The free and wall controls have broad lateral domains. The wall occupies $z\le1.05~\mu$m; the straight channel has a $2\times2~\mu$m accessible cross-section. These domains prescribe \emph{center} accessibility directly and are not eroded by a tracer radius a second time. There are $3\times3\times8000=72000$ histories.

\begin{table}[!htbp]
\centering\small
\setlength{\tabcolsep}{4pt}
\caption{Voxel control geometries ($L=3.5~\mu$m, $D=1~\mu$m$^2$/s, $T_c=20$ s): 8000 trajectories per condition. Error is relative to the analytical RMST for the sampled initial positions.}\label{tab:obstacles}
\begin{tabular}{@{}lccccl@{}}
\toprule
Control & $\Delta t$ (ms) & RMST (s) & Analytical (s) & Error [CI] (\%) & Decision \\ \midrule
free & 1.0000 & 1.5695 & 1.5109 & $3.88\ [2.03,5.72]$ & Inconclusive \\
free & 0.1250 & 1.5137 & 1.5109 & $0.19\ [-1.62,1.99]$ & Within \\
free & 0.0625 & 1.5239 & 1.5109 & $0.86\ [-1.01,2.73]$ & Within \\
Planar wall & 1.0000 & 2.8294 & 2.7356 & $3.43\ [1.39,5.46]$ & Inconclusive \\
Planar wall & 0.1250 & 2.7772 & 2.7356 & $1.52\ [-0.47,3.52]$ & Inconclusive \\
Planar wall & 0.0625 & 2.7859 & 2.7356 & $1.84\ [-0.17,3.84]$ & Inconclusive \\
Straight channel & 1.0000 & 1.6395 & 1.5109 & $8.51\ [6.58,10.44]$ & Outside \\
Straight channel & 0.1250 & 1.5619 & 1.5109 & $3.37\ [1.45,5.29]$ & Inconclusive \\
Straight channel & 0.0625 & 1.5242 & 1.5109 & $0.88\ [-0.96,2.71]$ & Within \\
\bottomrule
\end{tabular}
\end{table}

The straight channel identifies a clear discrepancy at 1 ms: $8.51\%$ RMST error, outside the $\pm3\%$ tolerance. At 0.0625 ms, the free and channel means meet the margin, while the wall remains inconclusive. No impossible lower-face escapes occur in the wall case. Small observed survival-curve discrepancies do not establish a uniform 2 pp error bound.

\clearpage
\subsection*{Continuous-detector controls}
These controls use $L=5~\mu$m, uniform initial $z\in[2,3]~\mu$m and $D=1~\mu$m$^2$/s. A solid wall at $z\le1.5~\mu$m with tracer radius $0.1~\mu$m puts its accessible boundary at $a=1.6~\mu$m. The channel half-width is $1~\mu$m. Each case has 8192 trajectories at each of two steps.

\begin{table}[!htbp]
\centering\small
\setlength{\tabcolsep}{4pt}
\caption{Continuous-detector controls: $L=5~\mu$m, $D=1~\mu$m$^2$/s, $T_c=100$ s, and 8192 trajectories per condition. No control was repeated at the two later refined steps.}\label{tab:continuous-controls}
\begin{tabular}{@{}lcccc@{}}
\toprule
Control & $\Delta t$ (ms) & RMST (s) & Analytical (s) & Error [CI] (\%) \\ \midrule
free & 0.125 & 3.0651 & 3.0840 & $-0.61\ [-2.40,1.17]$ \\
free & 0.0625 & 3.0557 & 3.0840 & $-0.92\ [-2.69,0.86]$ \\
wall & 0.125 & 5.4037 & 5.3336 & $1.31\ [-0.65,3.28]$ \\
wall & 0.0625 & 5.4004 & 5.3336 & $1.25\ [-0.72,3.22]$ \\
channel & 0.125 & 3.1090 & 3.0840 & $0.81\ [-1.01,2.63]$ \\
channel & 0.0625 & 3.0655 & 3.0840 & $-0.60\ [-2.40,1.20]$ \\
\bottomrule
\end{tabular}
\end{table}

\begin{figure}[!htbp]
\centering\includegraphics[width=.92\textwidth]{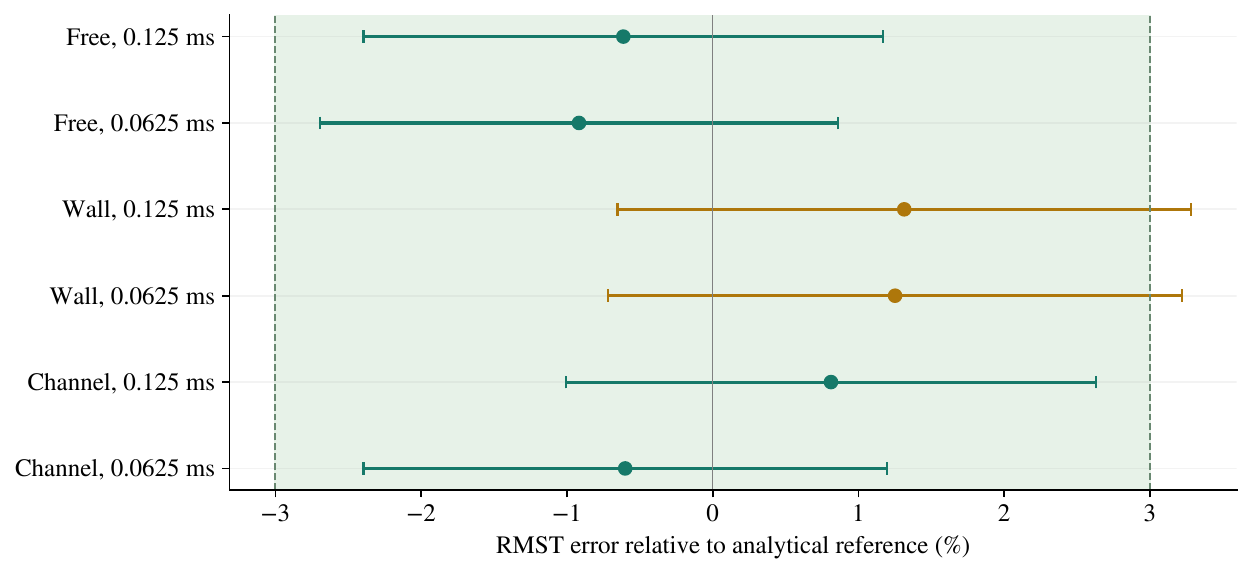}
\caption{Continuous-detector RMST errors and 95\% intervals relative to the analytical controls. Green intervals meet the 3\% margin; amber intervals are inconclusive.}
\end{figure}
The wall upper interval remains $3.22\%$ at 0.0625 ms. All six conservative uniform survival-error bounds exceed 2 pp, despite observed maxima of roughly 0.8--1.2 pp. Thus analytical verification is informative but not globally closed. No full-control suite was rerun at 0.03125 or 0.015625 ms; later steps retain the same detector and boundary rule.

\clearpage
\suppnote{Spatial geometry and transport comparison}
For a cell $C$ and solid set $\mathcal S$, the conservative admission rule is
\begin{equation}
\min_{\mathbf x\in C,\mathbf y\in\mathcal S}
\|\mathbf x-\mathbf y\|\ge a+10^{-12}~\mu\mathrm{m}.
\end{equation}
A partially invalid cell is excluded completely. Refinement can recover valid parts without changing the underlying solid. The tested resolutions have $h_{xy}=11.621$ or $5.811$ nm and $h_z=6.25$ or $3.125$ nm; all span the same local ROI and 50 physical layers.

\begin{table}[!htbp]
\centering\small
\setlength{\tabcolsep}{4pt}
\caption{Local safe-cell geometry. Accessible volume is normalized by the entire region of interest (ROI); through-connected volume is normalized by accessible volume. Exclusion is normalized by valid points in the fresh continuous-reference sample.}\label{tab:geometry}
\begin{tabular}{@{}ccccccc@{}}
\toprule
$d$ ($\mu$m) & $h_{xy}$ (nm) & $h_z$ (nm) & Accessible (\%) & Components & Through (\%) & Excluded (\%) \\ \midrule
0.20 & 11.621 & 6.250 & 53.782 & 120 & 99.995 & 3.312 \\
0.50 & 11.621 & 6.250 & 18.328 & 315 & 97.814 & 5.664 \\
0.20 & 11.621 & 3.125 & 54.304 & 152 & 99.995 & 2.364 \\
0.50 & 11.621 & 3.125 & 18.583 & 332 & 97.767 & 4.365 \\
0.20 & 5.811 & 6.250 & 54.203 & 299 & 99.995 & 2.555 \\
0.50 & 5.811 & 6.250 & 18.618 & 470 & 97.789 & 4.257 \\
0.20 & 5.811 & 3.125 & 54.721 & 296 & 99.995 & 1.634 \\
0.50 & 5.811 & 3.125 & 18.873 & 574 & 97.754 & 2.936 \\
\bottomrule
\end{tabular}
\end{table}

Six-face adjacency defines connectivity; edge/corner-only contacts do not connect components. A main component occupies at least 1\% of accessible volume. A through component touches both external $z$ faces. No $z$-face contact does not imply a completely enclosed physical cavity, because the ROI has artificial sides.

The final per-axis comparisons require: relative accessible-volume change below 3\%; unchanged identities/counts of main and main-through components; unchanged presence of a through path; through-volume-fraction change below 3 pp; and no observed invalid admissions. All four final diameter-by-axis comparisons meet these criteria. Main-component identities were checked with physically matched points, not counts alone. Total small-component counts still change.

\begin{figure}[!htbp]
\centering\includegraphics[width=\textwidth]{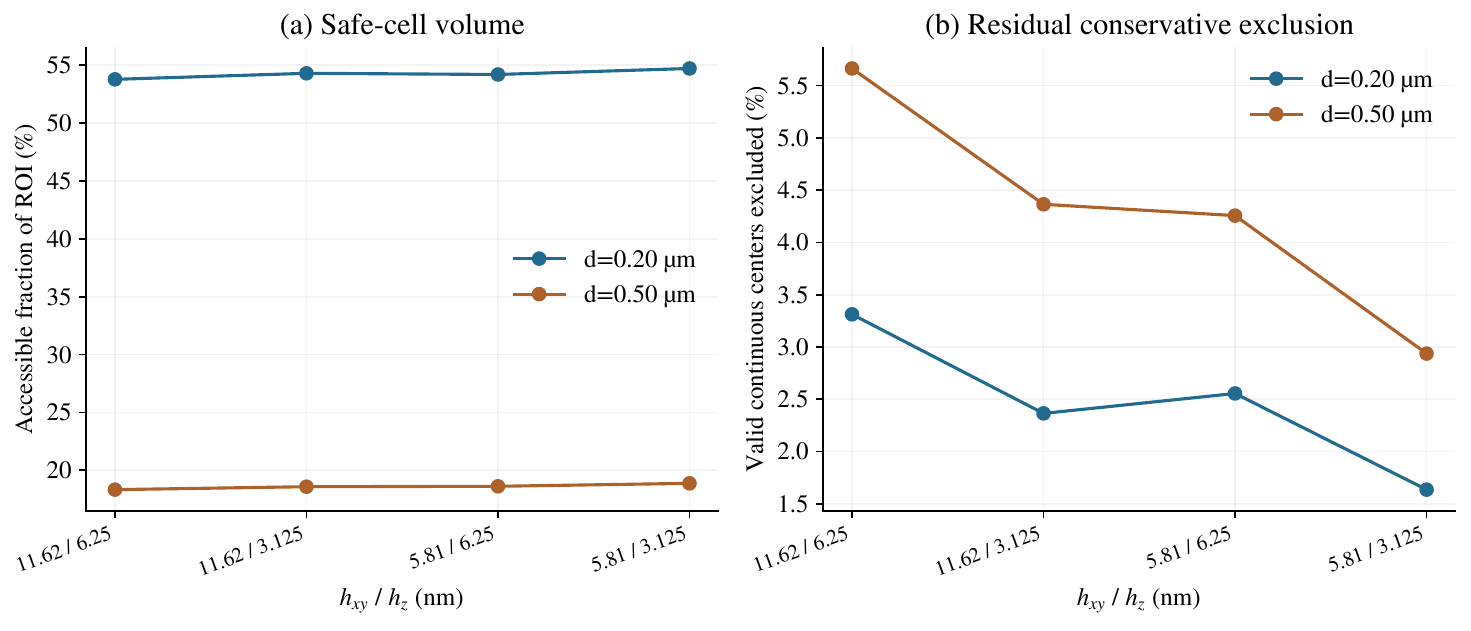}
\caption{Safe-cell volume and residual exclusion relative to a fresh continuous-reference sample of 200000 points. The cell volumes are exact raster counts; the exclusion estimates carry sampling uncertainty. The two axes have different denominators.}
\end{figure}

\clearpage
\subsection*{Residual error and kinetic comparison}
The finest safe-cell mask retains 54.7214\% of ROI volume for diameter $0.20~\mu$m and 18.8731\% for $0.50~\mu$m. The continuous-reference estimates are 55.5840\% and 19.4285\%, respectively. Among valid continuous-reference points, the fine grid excludes 1.634\% (95\% Wilson CI $[1.562,1.711]\%$) and 2.936\% ($[2.773,3.109]\%$). This is not the same denominator or criterion as a 3\% change between grids.

Checks included 76 block-labeling comparisons against full-volume labeling, conservation of parent safe cells under subdivision, 16384 direct cell--solid distance comparisons, and zero observed false admissions in 220000 reference positions per configuration and radius. These are finite checks under the idealized geometry, not anatomical error bounds.

A subsequent precision study compared transport on the finest safe-cell mask with the continuous detector at the same 0.125 ms step, and compared two continuous steps. It used 8192 new paired starts per diameter, conditioned to be valid in all three initial comparison masks.

\begin{table}[!htbp]
\centering\small
\setlength{\tabcolsep}{4pt}
\caption{Precision study at 100 s: 8192 paired trajectories per diameter and method. Cells denote the finest safe-cell mask at 0.125 ms; continuous geometry uses the same solid. None of these four comparisons establishes joint equivalence.}\label{tab:bridge}
\begin{tabular}{@{}clcc@{}}
\toprule
$d$ & Comparison (ms) & $\Delta P$ [CI] (pp) & $\Delta$RMST [CI] (\%) \\ \midrule
0.20 & Cells -- continuous & $-0.01\ [-0.04,0.01]$ & $3.97\ [1.57,6.36]$ \\
0.20 & 0.125 -- 0.0625 & $0.00\ [-0.04,0.04]$ & $0.62\ [-1.80,3.04]$ \\
0.50 & Cells -- continuous & $-1.29\ [-2.67,0.08]$ & $1.52\ [0.35,2.69]$ \\
0.50 & 0.125 -- 0.0625 & $-2.17\ [-3.57,-0.78]$ & $2.02\ [0.81,3.23]$ \\
\bottomrule
\end{tabular}
\end{table}

For the small tracer, the cell-versus-continuous RMST difference remains $3.97\%$ with a $[1.57,6.36]\%$ interval. For the large tracer, RMST meets the margin but escape remains inconclusive. The small-tracer temporal upper RMST limit is $3.037\%$, just above 3\%; it is not rounded into a pass.

Consequently, stability of local geometric metrics is not closure of spatial transport convergence. The voxel and continuous methods must not be described as interchangeable. The finest full-domain grid would require more than $10^{11}$ cells; the local reference was not a full-clot computation.

\clearpage
\suppnote{Nested temporal refinement of continuous transport}
The reference geometry is frozen at mechanical time 10 s; transport is followed for 100 s. The diameter is $0.50~\mu$m, $D=0.4~\mu$m$^2$/s, and the cohort contains 8192 initial centers. They were sampled from a disk of radius $1.19~\mu$m in $xy$ and $z\in[2,3]~\mu$m, conditioned on the common safe-mask region. Isolated components were not deliberately removed.

The original 0.0625 ms increment stream is reproduced in its saved order. A Brownian bridge creates 0.03125 ms proposals, and a second independent bridge creates 0.015625 ms proposals \citep{iowa_bb}. For a parent increment $\delta$ of duration $h$,
\begin{equation}
\delta_1=\delta/2+\eta,\qquad
\delta_2=\delta-\delta_1,\qquad
\eta\sim\mathcal N(0,Dh/2)
\end{equation}
in each coordinate. Each child has variance $Dh=2D(h/2)$ and the children sum to the parent. The bridge applies to free proposals, not already rejected positions; accepted paths can diverge after different collisions.

The first refinement reproduced 16 full 0.0625 ms histories exactly; the second reproduced 16 full 0.03125 ms histories. The final nested-noise check used 300000 scalar parent splits, verified quarter-step variances and correlations, and passed 20000 unchanged-detector comparisons. Production-only fine histories matched their joint-replay equivalents.

\begin{table}[!htbp]
\centering\small
\setlength{\tabcolsep}{4pt}
\caption{Continuous-geometry results at 100 s for a 0.50 $\mu$m-diameter tracer. There are 8192 trajectories at each step; the same initial centers are paired across methods.}\label{tab:final-summary}
\begin{tabular}{@{}ccccc@{}}
\toprule
$\Delta t$ (ms) & Escape (\%) & $S(100)$ & RMST [CI] (s) & Censored \\ \midrule
0.0625 & 53.345 & 0.46655 & $71.573\ [70.871,72.276]$ & 3822/8192 \\
0.03125 & 54.797 & 0.45203 & $70.732\ [70.026,71.438]$ & 3703/8192 \\
0.015625 & 53.650 & 0.46350 & $71.437\ [70.728,72.146]$ & 3797/8192 \\
\bottomrule
\end{tabular}
\end{table}

The final observed escape probabilities, 53.34\%, 54.80\%, and 53.65\%, are not monotonic with refinement. This prevents treating the last change as a continuing one-direction trend; it does not prove that all remaining differences are sampling noise. At the finest step, 3797 histories remain censored and RMST is 71.4368 s.

\begin{table}[!htbp]
\centering\small
\setlength{\tabcolsep}{4pt}
\caption{Final comparison: 0.03125 minus 0.015625 ms. The 100 s horizon is primary; shorter horizons are secondary. ``Within'' requires the entire interval to lie inside the prespecified margin.}\label{tab:final-comparison}
\begin{tabular}{@{}cccl@{}}
\toprule
$T_c$ (s) & $\Delta P$ [CI] (pp) & $\Delta$RMST [CI] (\%) & Escape / RMST \\ \midrule
10.0 & $-0.354\ [-0.722,0.014]$ & $0.076\ [-0.047,0.199]$ & Within / Within \\
20.0 & $-0.476\ [-1.223,0.271]$ & $0.124\ [-0.177,0.425]$ & Within / Within \\
50.0 & $1.245\ [0.042,2.448]$ & $-0.044\ [-0.761,0.673]$ & Inconclusive / Within \\
100.0 & $1.147\ [-0.194,2.489]$ & $-0.987\ [-2.154,0.180]$ & Inconclusive / Within \\
\bottomrule
\end{tabular}
\end{table}

\clearpage
\begin{figure}[!htbp]
\centering\includegraphics[width=.95\textwidth]{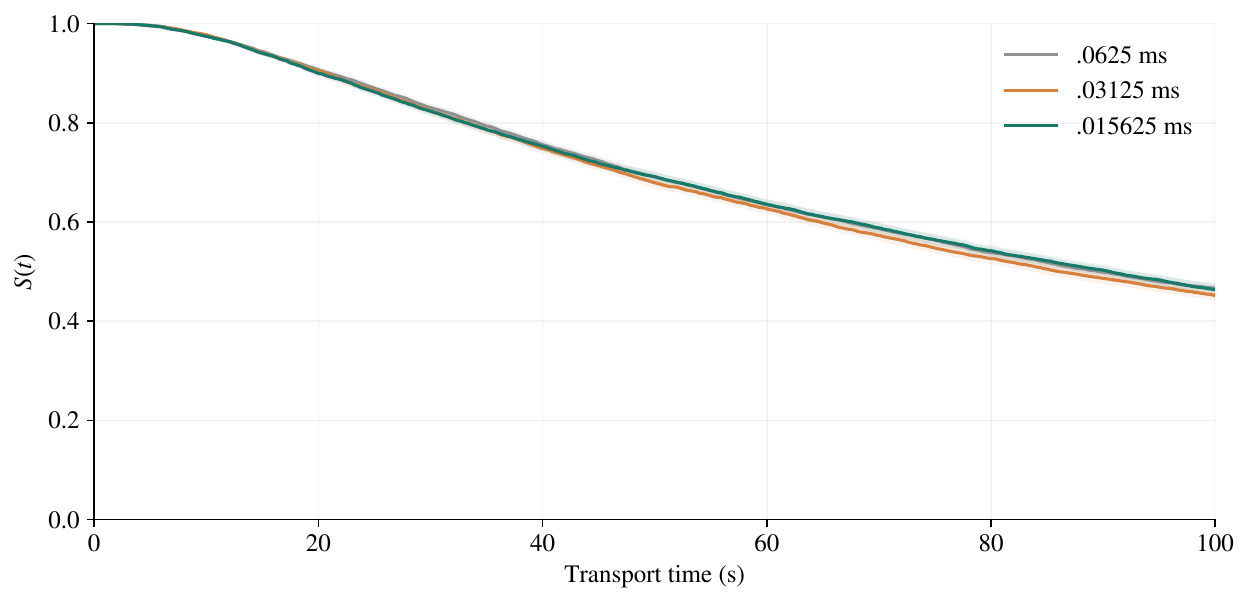}
\caption{Continuous-geometry survival at the three latest steps. Shading gives pointwise Wilson 95\% intervals. All curves use the same frozen local geometry, initial-center cohort, and physical diffusion coefficient.}
\end{figure}
\begin{figure}[!htbp]
\centering\includegraphics[width=\textwidth]{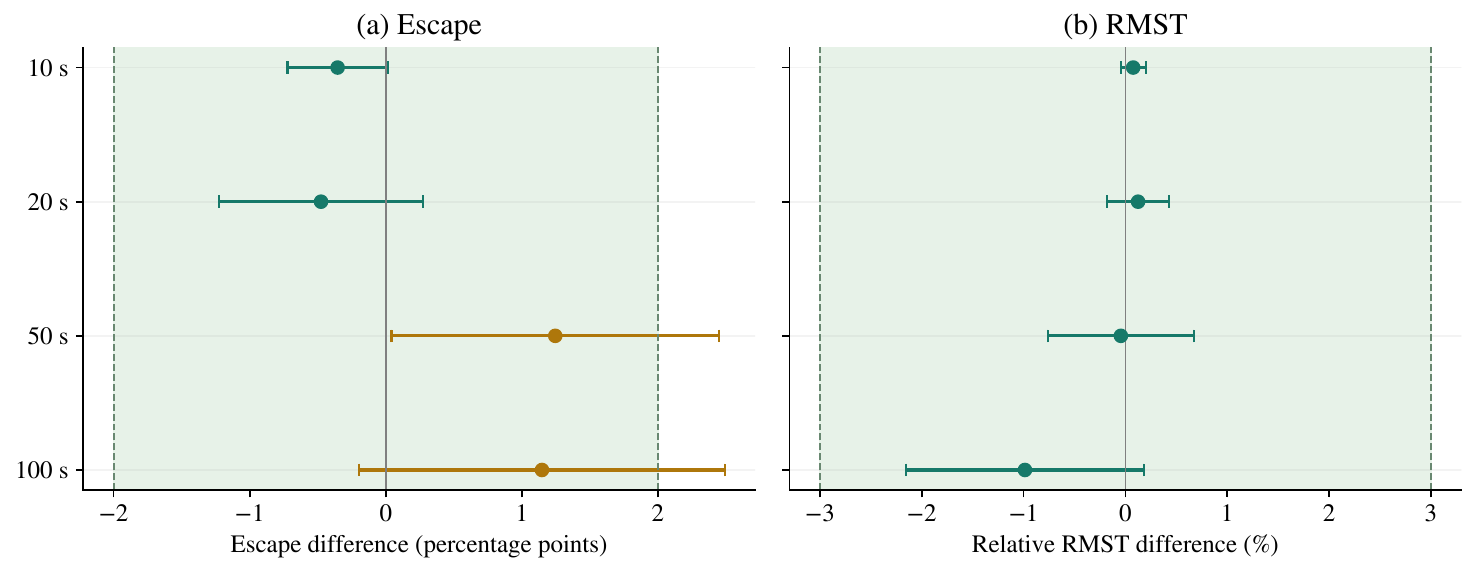}
\caption{Final paired temporal comparison. At 10 and 20 s both indicators meet their margins. At 50 and 100 s RMST meets its margin, but escape remains inconclusive. Green denotes within-margin intervals and amber denotes inconclusive intervals.}
\end{figure}
At 100 s the final escape interval is $[-0.194,2.489]$ pp and the relative RMST interval is $[-2.154,0.180]\%$. The phase therefore closes with RMST equivalence and unresolved joint equivalence, not a complete temporal-convergence claim. The earlier 0.0625/0.03125 ms comparison also remains inconclusive for escape.

All 32 batches were completed without outcome-dependent stopping or sample expansion. The final run took 147.32 min with four CPU processes. Its completion audit checked all 8192 new histories and reused records, source/data hashes, timing, censoring, counters, coordinate bounds, and exact empirical-survival integration. The 342 protected files were unchanged \emph{during that experiment}; subsequent authorized manuscript edits are separate from those archived audit snapshots.

\clearpage
\suppnote{Closure and reproducibility of the historical verification}
\textbf{Supported conclusions.} The specified collision and censoring rules pass their implementation checks. The fixed $H=70$ nm mask has a practically equivalent selected time-step pair. Local conservative geometric metrics meet their per-axis criteria. The final $H=100$ nm continuous comparison meets the restricted-mean tolerance for the tested large tracer and all four horizons.

\textbf{Unresolved items.} Joint escape/RMST equivalence at 50--100 s, uniform survival-curve equivalence, the mixed-boundary analytical tolerance, full cell-versus-continuous kinetic equivalence, complete-clot boundary handling, and transport with moving obstacles are not established. Neither $H=100$ nm nor the local ROI geometry is experimentally validated here.

\textbf{Operational choice.} Preserve the continuous detector and the finest tested 0.015625 ms calculation as a reference for this specific local problem. This is not an automatic change to production defaults or validation of the reflection protocol in Supplementary Note~\ref{note:current}. A larger or evolving system requires a documented treatment of its own boundaries and geometry. No new validation run was launched for this manuscript update.

\subsection*{Data inventory}
The companion directory \path{data/validation/} contains the source CSVs used to construct the tables and figures:
\begin{itemize}
\item \path{h70_comparison.csv}, \path{h70_summary.csv}, and \path{h70_additional_batches.csv}: combined fixed-mask results and batch sensitivity.
\item \path{obstacle_controls.csv} and \path{continuous_controls.csv}: analytical benchmark comparisons.
\item \path{spatial_geometry.csv} and \path{spatial_comparisons.csv}: conservative geometry and per-axis criteria.
\item \path{cell_continuous_comparisons.csv}: new-only precision comparison across geometry and time-step methods.
\item \path{continuous_dt31_comparisons.csv}: first nested refinement.
\item \path{continuous_final_comparisons.csv}, \path{continuous_final_summary.csv}, and \path{continuous_final_survival.csv}: final refinement, censoring, and survival curves.
\end{itemize}
\path{provenance.json} records each original source path and SHA-256 hash. Figure files are in \path{figures/validation/}. Underlying per-tracer histories, seeds, configurations, source files, and completion audits remain in their original experiment directories; these were not rerun or overwritten.

The $H=70$ nm study uses batch seeds 77000--77015. The $H=100$ nm precision cohort and both nested refinements use seeds 80600--80631, with particle identity preserved within each 256-history batch. Saved final coordinates are the last accepted interior positions, not exact face-intersection coordinates. Full intermediate tracer positions are not stored, although the fixed seeds and source versions reproduce proposals.

Mechanical replicas and Monte Carlo batches are distinct sampling units. More tracers improve precision conditional on a fixed network but do not estimate uncertainty across clots. Differences between two size/diffusion scenarios are not exclusively steric, and finite-time survival alone cannot establish binding, irreversible capture, permeability, or an independent aggregation rate.

\clearpage
\suppnote{Current full-common-domain reflection and homogeneous cohort}\label{note:current}
\subsection*{Distinct protocol and implementation checks}
The main-text cohort uses 50 layers with $H=100$ nm, a $0.25~\mu$m-radius tracer, $D=0.4~\mu$m$^2$/s, and continuous extrusions of backbone segments, cohesive links, and platelet disks. The lateral region is the intersection of the layer interiors on their native raster, not the central ROI of Supplementary Notes~S1--S7. Its boundary reflects tracer centers directly. Both $z=0$ and $z=5~\mu$m absorb the center. Central proposals have radius $1.19~\mu$m and $z\in[2,3]~\mu$m and are accepted by clearance, without connectivity conditioning.

The remaining Gaussian displacement is geometrically reflected at the first contact. Multiple active normals are orthogonalized; their normal subspace is reflected. A contact tied with an axial exit is processed first. If 256 contacts exhaust a local budget, the same increment is subdivided geometrically, retaining noise and physical duration. This resolves a verified 713-bounce narrow-gap path; it is not Brownian time-step refinement. Numerical failures terminate a run rather than produce censored particles. The method is not the force-dependent propagator of \citet{behringer2012}.

The reflection implementation passed 26510 test cases: 1500 analytic sphere contacts, 1500 cylinder contacts, 500 extruded-capsule checks against independent minimization, 20000 planar-image checks, 2000 lateral segment--box comparisons, 1000 multihit reversibility cases, and ten boundary-ordering, corner, and subdivision controls. These counts describe tests, not independent mechanical samples. A 4096-tracer straight-channel benchmark reduced the relative point error in RMST from 5.28\% at 0.5 ms to 1.17\% at 0.015625 ms, relative to the truncated analytical reference. It does not establish uniform error bounds over curved clot geometries.

\subsection*{Initial long-horizon temporal comparison}
On replica 1, $f_0=2.2$ nN, mechanical time 10 s, 256 paired histories compare 0.03125 and 0.015625 ms until 100 s. At the coarse step, 197 escape and 59 are censored, with RMST 58.8524 s. At the fine step, 192 escape and 64 are censored, with RMST 59.0619 s. There are 59 discordant escape pairs. Table~\ref{tab:reflection-equivalence} preserves the inconclusive decisions.

\begin{table}[htbp]
\centering\small
\caption{Initial 256-history reflection comparison, coarse minus fine. Marginal 97.5\% intervals use a Bonferroni allocation for nominal joint 95\% coverage of the two endpoints; bootstrap coverage is approximate. Original margins and decisions are preserved.}\label{tab:reflection-equivalence}
\begin{tabular}{@{}lccc@{}}
\toprule
Observable & Difference [interval] & Margin & Decision \\ \midrule
$S(100)$ & $-1.953\ [-12.238,8.408]$ pp & $\pm2$ pp & Inconclusive \\
Relative RMST & $-0.355\ [-6.865,6.650]\%$ & $\pm3\%$ & Inconclusive \\
\bottomrule
\end{tabular}
\end{table}
The survival interval uses conservative exact binomial bounds for paired discordance; RMST uses 20000 paired bootstrap resamples. Extending the horizon exposes late exits but does not increase the number of independent trajectories or establish equivalence. This initial comparison motivated the larger-sample study below. Spatial convergence of this complete-domain transport problem was not tested by either temporal comparison.

\subsection*{Larger-sample refinement and revised operational criterion}
A fixed-allocation study uses 8192 paired histories in each of three frozen states: A, replica 1 at 0 s; B, replica 1 at $f_0=2.2$ nN and 10 s; and C, replica 9 at $f_0=8.8$ nN and 15 s. These states represent two mechanical seed groups, not three independent mechanical replicas. Each state is tested at 0.03125, 0.015625, and 0.0078125 ms with a 100 s horizon, totaling 73728 verification trajectories, separate from the main-text cohort. The geometries, initial release law, reflection rules, tracer radius, and diffusivity are unchanged. New initial-position and Brownian seeds are used; within each state, all steps share the same accepted starting positions and nested Gaussian increments. No mechanical evolution was repeated.

The primary contrast is 0.015625 minus 0.0078125 ms. Survival intervals retain the same conservative construction from exact binomial bounds on positive and negative discordances. Relative RMST intervals use 20000 paired percentile-bootstrap resamples, with the finer RMST as denominator. Marginal coverage is 97.5\%, with nominal joint 95\% coverage for the two endpoints within a comparison; bootstrap coverage is approximate. Simultaneous interval coverage across all states is not claimed. The allocation and original margins, $\pm2$ pp and $\pm3\%$, were fixed before the new outcomes were inspected; no interim scientific analysis or outcome-dependent sample enlargement occurred.

\begin{table}[!htbp]
\centering\small
\setlength{\tabcolsep}{4pt}
\caption{Independent larger-sample temporal comparison: 0.015625 minus 0.0078125 ms, with 8192 paired histories per state and 100 s observation. Intervals are marginal 97.5\% intervals. Decisions use the revised survival margin of $\pm2.5$ pp and unchanged RMST margin of $\pm3\%$. The survival margin was revised after examining the results; under the original $\pm2$ pp margin, state B remains inconclusive.}\label{tab:temporal-revised}
\begin{tabular}{@{}lccc@{}}
\toprule
State & $\Delta S$ [interval], pp & Relative RMST [interval], \% & Revised decision \\ \midrule
A: R1, $t_m=0$ s & $0.024\ [-1.398,1.447]$ & $0.338\ [-0.629,1.317]$ & Within margins \\
B: R1, 2.2 nN, 10 s & $0.806\ [-0.782,2.392]$ & $0.318\ [-0.682,1.339]$ & Within margins \\
C: R9, 8.8 nN, 15 s & $-0.049\ [-1.424,1.326]$ & $0.149\ [-1.067,1.385]$ & Within margins \\
\bottomrule
\end{tabular}
\end{table}

After reviewing the completed results, the operational survival tolerance was revised to $\pm2.5$ pp, retaining $\pm3\%$ for relative RMST. Under this revised criterion, both indicators meet their margins in all three states (Table~\ref{tab:temporal-revised}). Under the original survival margin, state B remains inconclusive because its interval reaches $2.392$ pp. The revised acceptance is therefore a post-analysis tolerance assessment, not confirmation of the original criterion or a biologically calibrated error threshold \citep{lakens2017}. The intervals and raw outcomes are unchanged. At a survival level of 25\%, 2.5 pp corresponds to a 10\% relative difference, illustrating why the absolute tolerance must remain explicit.

The secondary 0.03125-minus-0.015625 ms contrasts meet even the original margins in all three states. This does not imply exact convergence or monotonic sample differences with refinement. All 1152 blocks finished without numerical failure. A legacy-noise build reproduced eight historical trajectories exactly in every recorded field; 4096 free-space paths per level verified nested increments, with maximum endpoint disagreement below $2.8\times10^{-11}~\mu$m. The selected 0.015625 ms step is retained under the revised practical criterion. This conclusion is restricted to the tested states and parameters and does not certify all 117 production geometries, uniform equivalence of $S(t)$, moving obstacles, or a revised stereological construction. Original and revised result summaries and the original protocol are retained separately in \path{data/temporal_confirmation_20261009/}.

\subsection*{Independent connectivity certification}
An auxiliary $512\times512\times100$ grid has $h_{xy}=92.96875$ nm and $h_z=50$ nm. For center-to-solid distance $d_c$ and cell half-diagonal $q$, the 1-Lipschitz distance bound gives
\begin{equation}
d_c-q\ge a+\epsilon\quad\hbox{(entirely valid)},\qquad
d_c+q\ge a-\epsilon\quad\hbox{(possibly valid)},
\end{equation}
where $\epsilon=10^{-12}~\mu$m. The inner graph uses six face neighbors, and the outer possible graph uses 26 neighbors to avoid ruling out corner-connected candidates. Both are clipped to the same common lateral region. Interior paths touching a $z$ face certify an exit connection. Up to 32 nearby exit-connected inner-cell centers within $0.5~\mu$m are also tried as endpoints of exact straight clearance-tested bridges. Failure of this bounded search is unresolved, not disconnection. Disconnection would require absence of a path even in the outer graph.

Among the 29952 unique histories, 20052 starts are independently connected, six disconnected, and 9894 unresolved. Replica 4 at 10 s has one disconnected start at 4.4 nN and two at 8.8 nN, with all remaining starts unresolved; the corresponding bounds are $[0,255/256]$ and $[0,254/256]$. At 15 s, replica 4 has one disconnected start at 2.2 nN and two at 4.4 nN. All six disconnected state-specific starts remain censored at 100 s. They involve two distinct physical coordinates reused across geometries, not six distinct pores. This is a small non-vacuous consistency check, not a general validation. The upper connectivity bound remains 100\% in states without a certified disconnected start. Replica 1 at 10 s and 4.4 or 8.8 nN has a lower bound of zero with all 256 starts unresolved, despite observed exits. Of the 6100 censored histories, 3380 start at certified connected centers, six at disconnected centers, and 2714 at unresolved centers. Geometry-classification files are preserved before the escape outcomes are used.

\subsection*{Homogeneous analysis, pairing, and provenance}
The cohort contains nine paired mechanical seed groups, four baselines, and frozen states at 0, 5, 10, and 15 s. Shared initial states reduce 144 labels to 117 geometries. The allocation is fixed at 256 histories per unique geometry. The completed six-block data at 0, 5, and 10 s supply 13824 histories unchanged, including the 256 compatible histories reused in the initial two-block pilot. Blocks 7--9 at those times add 6912 histories, and all nine blocks at 15 s add 9216, giving 29952 without double-counting shared initial states. Blocks 3--9 use consecutive groups of 50 layer seeds from 4100--4149 through 4400--4449. There are 36 force-specific stacks and 1800 layer evolutions, but only nine paired seed groups for between-replica comparisons.

The mechanical evolutions continue from their saved 10 s states to 15 s, preserving existing cohesive topology, link reference lengths, and stochastic continuation state. Original 0 s areas remain the contraction reference. The allocation was fixed before analyzing each batch, rather than all nine blocks being declared before the first pilot. Original batch protocols and execution records are retained separately. Exactly two fine-step histories contact the lateral wall: replica 8 at 2.2 nN and 10 s has one history with 981 contacts, and replica 8 at 8.8 nN and 15 s has one with 73 contacts. Thus there are 1054 lateral events in two histories, not 1054 independent particles. Neither occurs in the retained six-block subset. At the coarse step of the separate refinement comparison, one tracer also contacts the lateral wall.

Within each mechanical replica, Gaussian seeds are paired by particle index, but geometry-conditioned accepted initial positions can differ. Main-text contrast intervals use 10000 paired-index percentile bootstrap resamples \citep{efron1979}, with analysis seed 20260919 and replica draw order 1--9. This retains the original draws, metrics, and contrast intervals for replicas 1--6 at 0, 5, and 10 s. The same replica-specific index matrix is also used for 15-minus-10 s comparisons. Intervals quantify approximate conditional tracer-sampling variation, not between-network uncertainty, all initial-proposal dependence, or numerical bias. They are marginal rather than simultaneous. Supplementary CSVs retain both same-index initial matches and common physical positions independent of index. Force-control pairs share 26--225 physical coordinates among 256 starts, with at most 68 also matched by index.

The directory \path{data/homogeneous_pilot/} contains state metrics, geometry metrics, paired contrasts, survival horizons, connectivity counts, merged event histories, and per-start connectivity/escape classifications. Its provenance file records source paths and SHA256 hashes. The geometry and transport sources remain distinct from \path{data/validation/}, which stores the historical studies. No obsolete endpoint, rejection-only, or cohesion-omitting trajectories enter the homogeneous tables.

\textbf{Scope at closure.} The present result is a reproducible, frozen-state descriptive study. Nine mechanical replicas expose between-realization heterogeneity but do not establish population robustness; unresolved connectivity is not physical isolation; and no external structural or mechanical validation has been completed. The selected time step meets revised practical tolerances in three tested states, including one at 15 s; the original survival tolerance remains inconclusive in one state. Complete temporal/spatial convergence over the production cohort is not claimed. Further replicas, stronger connectivity certificates, and moving-network transport remain proposed continuations.

\end{document}